\documentclass[twocolumn]{aastex62}
\usepackage{amsmath}
\usepackage{lipsum}
\usepackage{blindtext}
\usepackage{graphicx} 
\usepackage{changepage}
\usepackage{array}
\usepackage{multirow}
\usepackage{tabularx}

\begin{document}
\title{\textbf{The Physics of the Coronal Line Region for Galaxies in MaNGA}}
\shorttitle{Physics of the Coronal Line Region}
\shortauthors{Negus et al.}

\author{James Negus}
\email{$^{\star}$ james.negus@colorado.edu}
\affil{The University of Colorado Boulder, 2000 Colorado Avenue, Boulder, CO 80309, USA}

\author{Julia M. Comerford}
\affil{The University of Colorado Boulder, 2000 Colorado Avenue, Boulder, CO 80309, USA}

\author{Francisco M\"uller S\'anchez}
\affiliation{The University of Memphis, 3720 Alumni Avenue, Memphis, TN 38152, USA}

\author{Jorge K. Barrera-Ballesteros}
\affiliation{Instituto de Astronomía, Universidad Nacional Autónoma de
	México, A.P. 70-264, 04510 México, D.F., México}

\author{Niv Drory}
\affiliation{McDonald Observatory, The University of Texas at Austin, 1 University Station, Austin, TX 78712, USA}

\author{Sandro B. Rembold}
\affiliation{Departamento de Física, CCNE, Universidade Federal de Santa Maria, 97105-900, Santa Maria, RS, Brazil}
\affiliation{Laboratório Interinstitucional de e-Astronomia - LIneA, Rua Gal. José Cristino 77, Rio de Janeiro, RJ - 20921-400, Brazil}

\author{Rogemar A. Riffel}
\affiliation{Departamento de Física, CCNE, Universidade Federal de Santa Maria, 97105-900, Santa Maria, RS, Brazil}
\affiliation{Laboratório Interinstitucional de e-Astronomia - LIneA, Rua Gal. José Cristino 77, Rio de Janeiro, RJ - 20921-400, Brazil}



\begin{abstract}
The fundamental nature and extent of the coronal line region (CLR), which may serve as a vital tracer for Active Galactic Nucleus (AGN) activity, remain unresolved. Previous studies suggest that the CLR is produced by AGN-driven outflows and occupies a distinct region between the broad line region and the narrow line region, which places it tens to hundreds of parsecs from the galactic center. Here, we investigate 10 coronal line (CL; ionization potential $\gtrsim$ 100 eV) emitting galaxies from the SDSS-IV MaNGA catalog with emission from one or more CLs detected at $\ge$ 5$\sigma$ above the continuum in at least 10 spaxels - the largest such MaNGA catalog.  We find that the CLR is far more extended, reaching out to 1.3 - 23 kpc from the galactic center. We cross-match our sample of 10 CL galaxies with the largest existing MaNGA AGN catalog and identify 7 in it; two of the remaining three are galaxy mergers and the final one is an AGN candidate. Further, we measure the average CLR electron temperatures to range between 12,331 K - 22,530 K, slightly above the typical threshold for pure AGN photoionization ($\sim$ 20,000 K) and indicative of shocks (e.g., merger-induced or from supernova remnants) in the CLR. We reason that ionizing photons emitted by the central continuum source (i.e. AGN photoionization) primarily generate the CLs, and that energetic shocks are an additional ionization mechanism that likely produce the most extended CLRs we measure. 

%


\end{abstract}

\keywords{Active galactic nuclei (16), Photoionization(2060), Emission line galaxies (459).}

\section{Introduction} \label{sec:intro}
An Active Galactic Nucleus (AGN), fueled by accretion onto its central supermassive black hole (SMBH; $M_{\rm{BH}} > 10^{6} M_{\odot}$), is one of the most energetic phenomena in the observable universe; it can generate a bolometric luminosity $ > 10^{46}$ erg s$^{-1}$, which can outshine the collective light of an entire host galaxy. The most powerful can also heat and photoionize gas tens of kpc away (e.g., \citealt{1981ApJ...247..403H,1987ApJ...321L..29M,2012MNRAS.420..878K,2013MNRAS.430.2327L,2017ApJ...849..102C}).


Moreover, feedback - the mechanism by which AGN processes (e.g., photoionization, radio jets, and winds) impact nearby matter - has been shown to influence the evolution of the host galaxy  (e.g., \citealt{2012ARA&A..50..455F, 2015MNRAS.452.2553J, 2017MNRAS.471...28K, 2017ApJ...843...98R}). On one hand, AGN feedback may quench star formation and limit the host galaxy's growth (e.g., \citealt{2008ApJS..175..390H}). On the other, AGN feedback can impart extreme pressure on the neighboring interstellar medium and generate high density conditions favorable for star formation (e.g., \citealt{2005MNRAS.364.1337S}). 

To unravel the dynamics of AGN-galactic co-evolution, reliable AGN identification is an essential first step. To address this, we consider the Unified Model of AGNs \citep{1993ARA&A..31..473A}. In this framework, an AGN is classified as either Type I or Type II. Type I are viewed pole-on and are observed to have broad (FWHM $>$ 1,000 km s$^{-1}$) and narrow (FWHM $<$ 1,000 km s$^{-1}$) emission lines, whereas Type II  are viewed edge-on and are observed to only have narrow emission lines. The physical structures that correspond to these regions are the broad line region (BLR) and the narrow line region (NLR), respectively. 

The NLR is the largest observable structure in the UV, optical, and near-IR that is directly impacted by the ionizing radiation from an AGN. Measurements from a \textit{Hubble Space Telescope} snapshot survey, in addition to ground based [OIII] observations, suggest this region extends hundreds of pc to several kpc from the nuclear core (e.g., \citealt{1996ApJS..102..309M, 2003ApJS..148..327S, 2011ApJ...739...69M}). However, the NLR surrounding an AGN may feature active star formation, which can also produce narrow lines (e.g., [NII] $\lambda6584$ and [OIII] $\lambda5007$; see \citealt{2016A&A...590A..37S}). As a result, the NLR is not always an ideal tracer for exclusive AGN activity. 

Comparatively, the BLR resides much closer to the SMBH (within $\sim$ 0.1 kpc; e.g., \citealt{2004ASPC..311..169L, 2011A&A...525L...8C}). The enhanced cloud velocities found in this region, near the accreting SMBH, provide reliable signatures for nuclear AGN activity. However, the BLR is not spatially resolved in most existing spectroscopic surveys. Further, for Type II AGN, the BLR is often obscured by a dusty torus that can attenuate optical and UV emission (e.g., \citealt{2019MNRAS.tmp.2203H}). 

For most AGNs, though, highly ionized species of gas with ionization potentials (IPs) $\gtrsim$ 100 eV (termed ``coronal lines"; CLs; e.g., [FeVII] and [NeV]) are proposed to exist in the nuclear region between the compact BLR  and the more extensive NLR (e.g., \citealt{1943ApJ....97...28S, 2016ApJ...824...34G}). These lines require extremely high energies for production, above the limit of stellar emission (55 eV; \citealt{2001ApJ...549L.151H}), likely trace the strong ionizing continuum of an AGN \citep{1943ApJ....97...28S}, and have been linked to AGN-driven outflows (e.g., \citealt{2011ApJ...739...69M, 2013MNRAS.430.2411M, 2020ApJ...895L...9R, 2021MNRAS.500.2666C}). As a result, detecting CLs may help identify AGNs and AGN-driven outflows in large spectroscopic galactic surveys, which are essential for understanding the influence of AGN feedback on the host galaxy.


CLs have been observed in the spectra of some AGNs; however, their nature and physical extent are still not very well understood (e.g., \citealt{1984MNRAS.208..347P, 1994A&A...288..457O, 2005MNRAS.364L..28P, 2011ApJ...739...69M}). 
Prior studies (e.g., \citealt{2011ApJ...743..100R}) suggest that they reside in a distinct zone of extremely hot and collisionally ionized plasma between the BLR and the NLR, in the coronal line region (CLR). Further, \cite{2010MNRAS.405.1315M} analyzed the structure and physical properties of the CLR for 10 pre-selected AGNs. They used the  \textit{Hubble Space Telescope}/Space Telescope Imaging Spectrograph to study [NeV] $\lambda$3427, [FeVII] $\lambda$3586, $\lambda$3760, $\lambda$6086, [FeX] $\lambda$6374, [FeXIV] $\lambda$5303, [FeXI] $\lambda7892$, and [SXII] $\lambda$7611 emission in their sample. They determined that the CLR extends between 10 - 230 pc from the nuclear core (similar to \citealt{{2011ApJ...739...69M}} and \citealt{2021MNRAS.tmp..778R}). They also measured CLR electron temperatures to vary between 10,000 K - 20,000 K (below the pure AGN photoionization threshold of $\sim$ 20,000 K)  and compared the ionization properties of the CLR to varying photoionization and shock models (e.g., \citealt{1996A&A...312..365B,1996ApJS..102..161D}). They concluded that AGN photoionization from a central source is the sole physical mechanism producing the lines in their sample. Moreover, \cite{2009MNRAS.397..172G} scanned the sixth Sloan Digital Sky Survey (SDSS) data release (\citealt{2008ApJS..175..297A}) and investigated the CLR in 63 AGNs with [FeX] $\lambda$6374 emission (IP = 233.60 eV), without an a priori assumption that emission would only be found in AGNs. They also studied [FeXI] $\lambda$7892 (IP = 262.10 eV) and [FeVII] $\lambda$6086 (IP = 99.10 eV) emission in these AGNs. They determined that photoionization is also the source of the CLs (inferred using X-ray observations from \textit{Rosat}; \citealt{1999A&A...349..389V,2000IAUC.7432....3V}). The authors then measured the extent of the CL emitting clouds in their sample and found that the lines with higher IPs (e.g., [FeX] and [FeXI]) feature larger FWHM values, consistent with the lines occupying a region closest to the BLR, at a scale comparable to the dusty torus. They determined that the lower IP [FeVII]-emitting regions feature narrower lines and likely merge with the traditional NLR.

Here, we use the SDSS's Mapping Galaxies at Apache Point Observatory (MaNGA) integral field unit (IFU) catalog, from the survey's fifteenth data release in its the fourth phase (SDSS-IV), to further evaluate the ionization mechanism(s) of the CLR and to better understand the relationship between CLs and AGN activity. Large IFU spectroscopic surveys are critical for studying CLs because they provide spatially resolved galactic spectra that can trace the full extent of the CLR. Additionally, MaNGA has observed $\sim$ 10,000 nearby galaxies, and the eighth data release of the catalog contains 6,263 unique galaxies, making it one of the largest IFU survey of galaxies currently available. With our custom pipeline, we detect 10 galaxies with emission from one or more CLs detected at $\ge 5\sigma$ above the continuum in at least 10 spaxels ($\sim$ 0.16\% of the sample), which makes it the most extensive such catalog of MaNGA CL galaxies to date.

This paper is outlined as follows: Section \ref{sec:obs} details the technical components of the SDSS-IV MaNGA survey and its data pipeline, Section \ref{sec:data} describes the methodology we use to build the CL catalog and to analyze the physical properties of the CLR, Section \ref{sec:results} reviews our results,  Section \ref{sec:discussion} provides interpretations of our findings, and Section \ref{sec:conclusion} includes our conclusions and intended future work. All wavelengths are provided in vacuum and we assume a $\Lambda$CDM cosmology with the following values: $\Omega_{M} = 0.287$, $\Omega_{\Lambda} = 0.713$ and $H_{0} = 69.3$ $ \rm{km}$ $\rm{s}^{-1}$ $\rm{Mpc}^{-1}$. 
\vspace*{-3mm} 
\section{Observations} \label{sec:obs}
\subsection{The MaNGA Survey}
To conduct our analysis, we utilize the largest IFU spectroscopic survey of galaxies to date, the SDSS-IV MaNGA catalog \citep{2015ApJ...798....7B,2015AJ....149...77D,2016AJ....152...83L, 2016AJ....152..197Y, 2017AJ....154...28B, 2017AJ....154...86W}. MaNGA began taking data in 2014 using the SDSS 2.5 m telescope \citep{2006AJ....131.2332G}, and has logged the spectra for $\sim$ 10,000 nearby galaxies (0.01 $<$ \textit{z} $<$ 0.15; average \textit{z} $\approx$ 0.03) with stellar mass distributions between $10^{9}$ $M_{\odot}$ and $10^{12}$ $M_{\odot}$. The spectra were taken at wavelengths between 3622 \AA$\>$ - 10354 \AA, with a typical spectral resolution of $\sim$ 2000 and a velocity resolution of $\sim$ 60 km s$^{-1}$ (see \citealt{2015ApJ...798....7B}). 

MaNGA relies upon dithered observations using IFU fiber-bundles, assembled individually from the Baryon Oscillation Spectroscopic Survey Spectrograph \citep{2013AJ....146...32S}. The IFUs are grouped into hexagonal grids with field-of-view (FoV) diameters between 12.$^{\prime\prime}$5 to 32.$^{\prime\prime}$5 - each distribution is matched to the apparent size of the target galaxy \citep{2015ApJ...798....7B}. The exact configuration of the fiber-fed system consists of two 19-fiber IFUs (12.$^{\prime\prime}$5 FoV), four 37-fiber IFUs (17.$^{\prime\prime}$5 FoV), four 61-fiber IFUs (22.$^{\prime\prime}$5 FoV), two 91-fiber IFUs (27.$^{\prime\prime}$5 FoV), and five 127-fiber IFUs (32.$^{\prime\prime}$5 FoV). The resulting pluggable science and calibration IFUs generate galactic spectroscopic maps out to at least 1.5 times the effective radius; the typical galaxy is mapped out to a radius of 15 kpc. Each MaNGA spatial pixel, or spaxel, covers 0.$^{\prime\prime}$5 $\times$ 0.$^{\prime\prime}$5 and the average full-width half maximum (FWHM) of the on-sky point spread function (PSF) is 2.$^{\prime\prime}$5, which corresponds to a typical spatial resolution of 1 -2  kpc \citep{2015AJ....149...77D}.  


\subsection{MaNGA Data Reduction Pipeline}
The MaNGA Data Reduction Pipeline (DRP) generates flux calibrated and sky-subtracted data for each galaxy in a FITS file format that is used for scientific analysis \citep{2016AJ....152...83L}. The resulting DRP data product is run through MaNGA's Data Analysis Pipeline (DAP; \citealt{2019arXiv190100856W}) that provides spectral modeling and science data products, such as stellar kinematics (velocity and velocity dispersion), emission-line properties (kinematics, fluxes, and equivalent widths), and spectral indices (e.g., D4000 and Lick indices). The data products are publicly released periodically as MaNGA Product Launches (MPLs).  

To construct our catalog of CL galaxies in MaNGA, we use MaNGA's eighth data release (MPL-8), which contains data for 6,263 unique galaxies. 
\section{Analysis} \label{sec:data}
\subsection{Spectral Fitting}\label{subsec:specfit}
The MaNGA DAP provides fits for prominent emission lines (e.g., H$\alpha$, H$\beta$, and [OIII] $\lambda5007$). However, it does not offer fits for the CLs of interest in this paper (Table \ref{tab:coronal}), which we select due to their high ionization potentials ($\gtrsim$ 100 eV), their presence within the wavelength range of MaNGA, and for comparison with prior CLR studies that featured these lines (e.g., \citealt{1989ApJ...343..678K, 1991A&A...250...57A, 1994A&A...288..457O, 2006ApJ...653.1098R, 2009MNRAS.394L..16M, 2010MNRAS.405.1315M, 2015MNRAS.448.2900R}). As a result, we construct a custom pipeline to scan for [NeV] $\lambda 3347$, $\lambda 3427$, [FeVII] $\lambda 3586$, $\lambda 3760$, $\lambda 6086$, and [FeX] $\lambda 6374$ emission at $\ge 5 \sigma$ above the background continuum in the 6,623 galaxies in MPL-8. We use this high 5$\sigma$ threshold to ensure we identify unambiguous CL emission. 

We first access the DRP to extract the necessary data cubes for each MaNGA galaxy. The data cubes provide a spectrum for each individual spaxel across the FoV of each galaxy (spaxel arrays vary between 32 $\times$ 32 spaxels to 72 $\times$ 72 spaxels, depending on IFU configuration). We then use the spectroscopic redshifts of each galaxy, adopted from the NASA Sloan Atlas catalogs using single-fiber measurements \citep{2011AJ....142...31B}, to shift the spectra to rest vacuum wavelengths and to impose a minimum redshift threshold ($\textit{z}_{\rm{min}}$) for CLs near the lower wavelength limit of MaNGA (3622 \AA; Table \ref{tab:redshift}). In some instances, CL rest wavelengths are redshifted out of MaNGA's spectral range.     

\begin{figure}[t]
	\includegraphics[height = 2.1in]{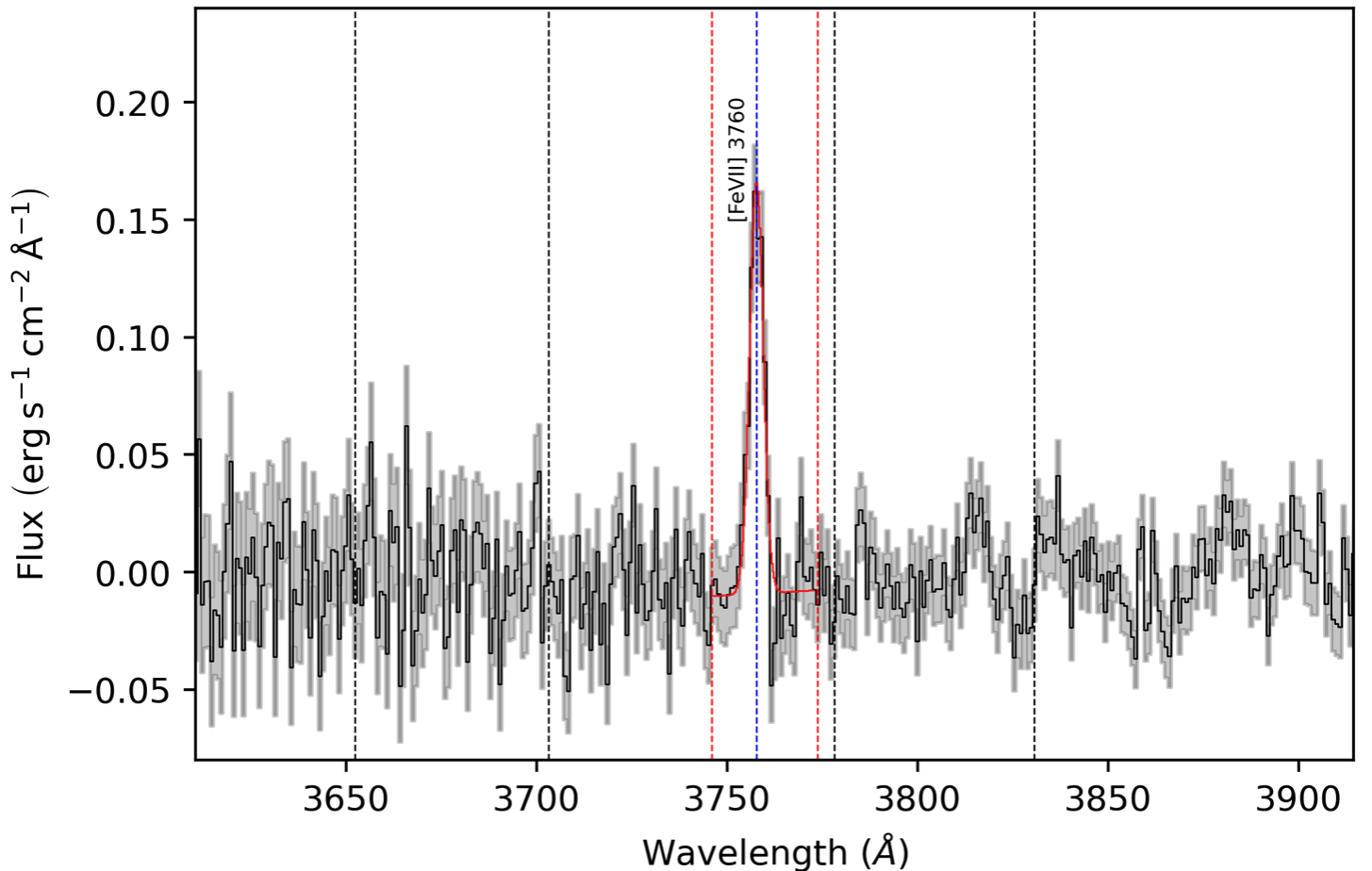}
	\caption{A sample spectrum from an individual spaxel showing the [FeVII] $\lambda3760$ line detected at $\ge 5 \sigma$ above the continuum in J1349. The solid black line is the continuum subtracted spectrum, the shaded gray region is the uncertainty, the solid red line represents the best fit, the red dotted vertical lines mark the fitting window, the blue dotted line signifies the rest wavelength of the [FeVII] $\lambda3760$ line, and the two sets of black  dotted vertical lines correspond to the neighboring continuum windows where the RMS flux values of the continuum are calculated.}
	\label{fig:spec1}
\end{figure}

	
%
Once the spectra are wavelength corrected, we analyze each individual spaxel for each galaxy and perform a polynomial fit on a narrow spectral region, $\sim$ 300 \AA $\>$ wide, of continuum near the CL to model the background stellar continuum and subtract it. We then attempt a single Gaussian fit on a $\sim 30$ \AA $\>$ region centered on the rest wavelengths of the CLs. This wavelength range was shown to sufficiently capture the full extent of CL emission in our preliminary scans, even for CLs with FWHM $>$ 1,000 km s$^{-1}$. We subsequently determine the root mean square (RMS) flux of two continuum regions ($\sim 60$ \AA $\>$wide) that neighbor each target CL, free of absorption or emission lines, and require that CL amplitudes are detected at $\ge 5\sigma$ above the mean RMS flux values in these continuum regions. We consider the spectral resolution of MaNGA (R = $\lambda$/$\Delta \lambda$ $\sim$ 1400 at 3600 \AA; R $\sim$ 2000 at 6000 \AA; \citealt{2013AJ....146...32S}) to eliminate fits with $\Delta \lambda$  $<$ 2.4 \AA $\>$ (for [NeV] $\lambda$3347, 3427), $<$ 2.6 \AA $\>$ (for [FeVII] $\lambda$3586, 3760), and $<$ 3 \AA $\>$ (for [FeVII] $\lambda$6086 and [FeX] $\lambda 6374$; we do not detect the [FeX] $\lambda 6374$ line using this criteria). We also require that each galaxy in our catalog features $>$ 10 CL-emitting spaxels to ensure we are detecting sources with definitive CL emission. We show an example of a single Gaussian fit for the [FeVII] $\lambda3760$ line in Figure \ref{fig:spec1}.

\begin{table}[t]
	\renewcommand{\thetable}{\arabic{table}}
	\centering
	\caption{Target CLs} 
	\begin{tabular}{cccc}
		\hline
		\hline
		Emission Line &  Wavelength &IP& \textit{z}$_{\rm{min}}$  \\ {} & {(\AA)} & (eV) & {} \\
		\hline
		\rm{[NeV]} &    3347 &   126.22 &  0.088\\
		\rm{[NeV]} &    3427 &   126.22 & 0.063\\
		\rm{[FeVII]} & 3586     & 125.0   & 0.016\\
		\rm{[FeVII]} &  3760   &  125.0   & -\\
		\rm{[FeVII]} &  6086   &   125.0  & -\\
		\rm{[FeX]} &  6374   &  262.1  & -\\
		\hline
		\multicolumn{4}{p{8cm}}
		{Note: Columns are (1) emission line, (2) rest wavelength, (3) ionization potential, and (4), minimum redshift value required for MaNGA detection.}
	\end{tabular}
	\label{tab:redshift}
	\label{tab:coronal}
\end{table}

\subsection{Coronal Line Flux Maps}\label{sec:fluxmaps}
\begin{figure}[t]
	\includegraphics[height = 2.8in]{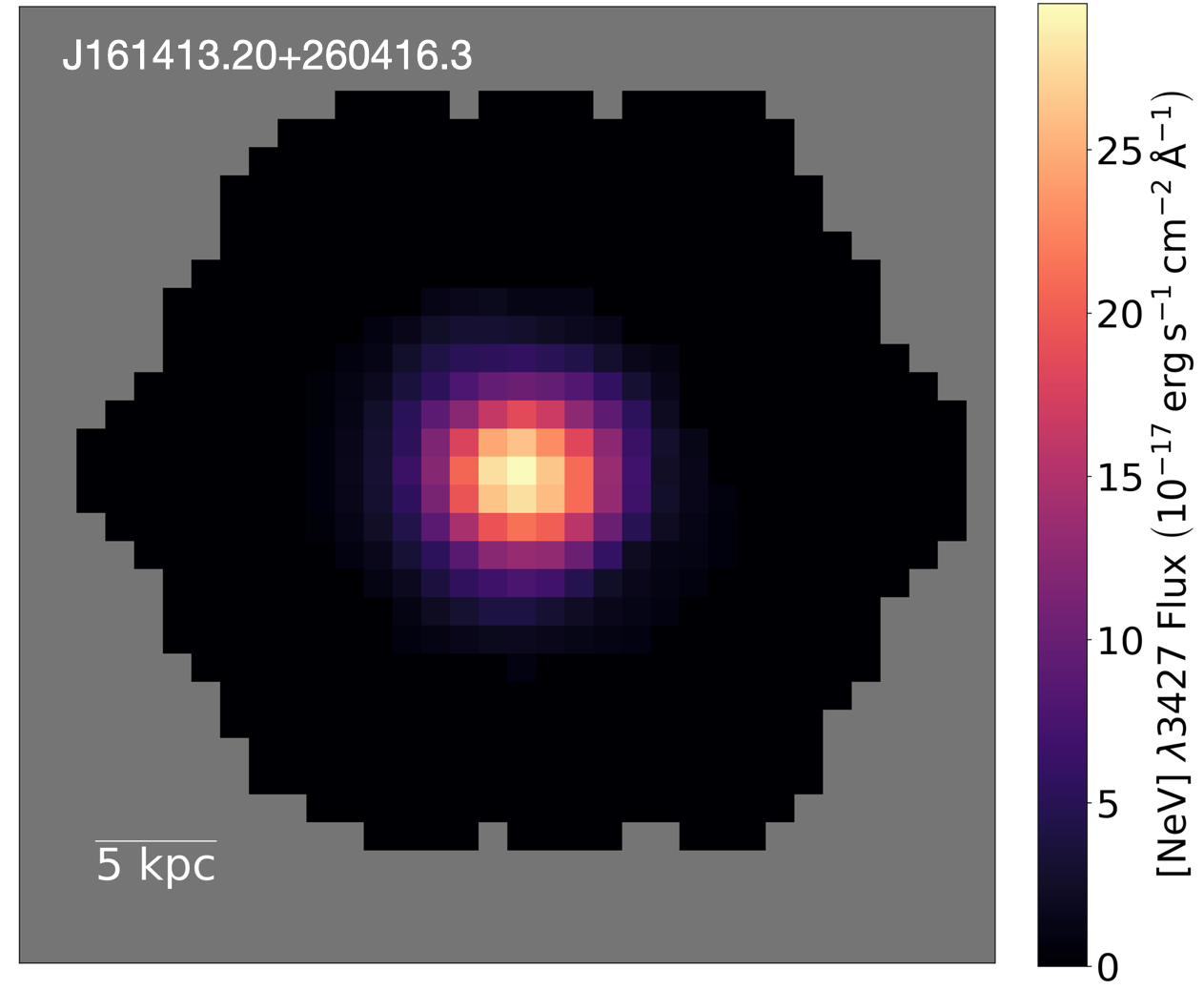}
	\caption{A sample CL flux map showing [NeV] $\lambda3427$ emission detected $\ge 5 \sigma$ above the continuum in J$1614$. For this galaxy, the strongest [NeV] $\lambda3427$ emission is located near the center of the galaxy. The gray region is outside of the MaNGA FoV.}
	\label{fig:examplefluxmap}
\end{figure}

We use flux maps to trace the strength and distribution of the CLs in the CLR. To generate these maps, we use the integrated CL flux value from each spaxel for each CL galaxy. We show an example flux map in Figure \ref{fig:examplefluxmap}.


For each MaNGA observation, the photometric center corresponds to the galactic center (\citealt{2016AJ....152..197Y}). We use this position and the galaxy's inclination angle to determine the de-projected galactocentric distance of each CL spaxel.  The MaNGA DAP provides the ratio of the semi-minor to semi-major axes (b/a) for each galaxy, and we use this value to determine the cosine of each galaxy's inclination angle (i): $\rm{cos}$($\rm{i}$)= b/a. The de-projected distance of each CL spaxel to the center of the galaxy is then measured by: 
\begin{equation}
\begin{aligned}
\label{eqn:1}
D_{\rm{Spaxel}}&= \sqrt{(x - x_{\rm{center}})^{2} + \big((y - y_{\rm{center}})*\rm{cos}(i) \big)^{2}}
\end{aligned}
\end{equation}
Next, we convert the spaxel distances to a physical unit (kpc) using the \texttt{astropy.cosmology} \texttt{Python} package. The resulting value corresponds to the distance of each CL emitting spaxel, and  the coronal line distance (CLD) corresponds to the distance of each galaxy's most extended CL-emitting spaxel (from the galactic center).
\subsection{Coronal Line Intensity Profiles}\label{sec:power-law}
\begin{figure}[t]
		\includegraphics[height = 3.4in]{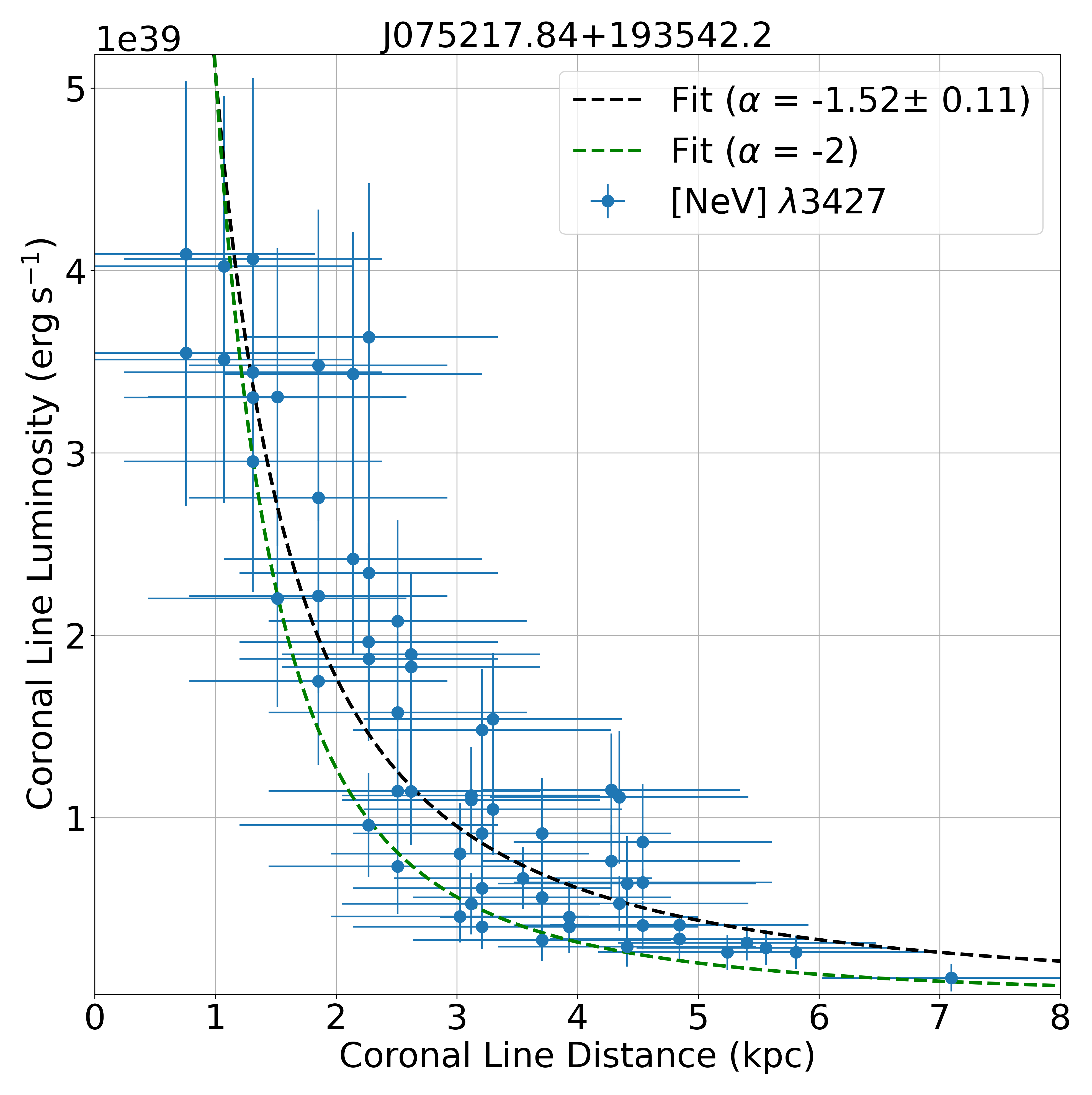}
		\caption{A sample plot of CL luminosity vs. de-projected galactocentric distance for the [NeV] $\lambda3427$ emitting spaxels in J$0752$. The black curve is the measured power-law fit for the data ($\alpha = -1.52$). The green curve is the model fit ($\alpha = -2$) for pure AGN photoionization.}
		\label{fig:plaw}
	\end{figure}

We explore CL intensity as a function of de-projected galactocentric distance to help determine the ionization mechanism(s) producing the CLs. We consider the study conducted by \cite{2012ApJ...747...61Y}, which analyzed ionization sources in galaxies with spatially extended emission line regions. The authors measured integrated flux profiles for a system of ionizing sources and found that the strength of ionizing flux, as a function of galactocentric distance, should decay with a power law index of $\alpha$ = -2 (i.e. obey the inverse square law) for pure AGN photoionization. They used this model to differentiate AGN photoionization from other sources (e.g., shocks). \cite{2002RMxAC..13..213T} also used this model to show that jet-induced shocks primarily ionize extended emission line regions along the axis of radio jets for several AGNs, and that AGN photoionization dominates within the nuclear regions. We fit a power law for each CL galaxy, and measure the best-fit power law index to help determine the ionization source(s) for the CLs. We present a sample power-law fit in Figure \ref{fig:plaw}. 

\subsection{Stellar Shock Diagnostics}
\label{shockssec}
We consider the role of supernova remnant (SNR) shocks in our analysis to determine their role in the production of CL emission. The first SNR identified in an external galaxy was observed by \cite{1972ApJ...178L.105M, 1973ApJ...180..725M}. The authors relied upon the strength of the [SII] $\lambda\lambda$6717, 6731 doublet with respect to the H$\alpha$ line to differentiate SNR shocks from photoionized regions. \cite{1978MmSAI..49..485D} and \cite{1980A&AS...40...67D} refined this relation empirically, and determined that regions with [SII] ($\lambda6717  + \lambda6731$)/H$\alpha$ $>$ 0.4 feature SNR shocks. The use of this threshold is widely adopted to identify SNRs (e.g., \citealt{1979ApJS...39....1R,1984ApJ...276..653D, 1995AJ....110..739L,2012ApJS..203....8B, 2014ApJ...786..130L}), and we employ it in our analysis. For each CL galaxy, we also divide the number of CL-emitting spaxels that feature SNRs by the total  number of CL-emitting spaxels. The resulting value is the fraction of SNRs in each galaxy's CLR (``SNR Frac'' in Table \ref{tab:cltable}). We calculate this parameter to determine if elevated fractions of SNRs correspond to enhanced CL production, which would suggest that SNR shocks play a major role in the production of CLs. The MaNGA DAP provides flux measurements for the [SII] $\lambda$6717, [SII] $\lambda$6731, and H$\alpha$ lines. 
	\vspace*{-3mm} 
\subsection{Electron Temperature and Electron Number Density Diagnostics}\label{sec:emlinediag}
\label{tempintro}
The electron temperatures and electron number densities of emission line regions provide direct insight into their ionization source(s). Specifically, temperatures between 10,000 K - 20,000 K are typical for photoionization equilibrium, as opposed to collisional shocks, for example, which can induce temperatures up to 10$^{6}$ K (e.g., \citealt{1984QJRAS..25....1O}). Furthermore, number densities on the order of $n_{\rm{e}}$ $\approx$ $10^{2}$ - $10^{4}$ cm$^{-3}$ are typical of the NLR; $n_{\rm{e}}$ $>$ 10$^{9}$ cm$^{-3}$ for the BLR (e.g., \citealt{1978ApJ...223...56K, 1997iagn.book.....P}).



We measure these parameters for the CL galaxies in our sample using \href{https://github.com/Morisset/PyNeb_devel/blob/master/docs/Notebooks/PyNeb_manual_7.ipynb}{Pyneb}, a \texttt{Python} package that evolved from the \texttt{IRAF} package \texttt{Nebular} \citep{1995PASP..107..896S, 1998ASPC..145..192S, 2015A&A...573A..42L}. \texttt{Pyneb's} \textbf{getCrossTemDen} routine iterates over the temperature sensitive flux ratio [OIII] ($\lambda$5007 + $\lambda$4959)/$\lambda$4363 and
the density sensitive flux ratio [SII] $\lambda$6717/ $\lambda$6731 to provide self-consistent temperature and density values (see \citealt{2006agna.book.....O}). For intermediate densities (n$_{\rm{e}}$ $\approx$ 10$^{3}$ cm$^{-3}$), which are found beyond the BLR (e.g., \citealt{1990agn..conf.....B}):
\begin{equation}
	\begin{aligned}
		\label{eqn:2}
		\frac{\rm{j}_{\lambda6717}}{\rm{j}_{\lambda6731}} \propto \frac{\rm{n}_{\rm{e}}}{\rm{n}_{\rm{e}}^{2}} \propto \rm{n}_{\rm{e}}^{-1}
	\end{aligned}
\end{equation} 
where j is the emissivity of each line and n$_{e}$ is the electron number density. For the temperature sensitive ratio (e.g., \citealt{1997iagn.book.....P}):
\begin{equation}
	\begin{aligned}
		\label{eqn:3}
	\frac{\rm{F}(\lambda4959 + \lambda5007)}{\rm{F}(\lambda4363)} \approx \frac{7.33 \> \rm{exp}\>(3.29 \times 10^{4}/T_{e})}{1 + 4.5 \times 10^{-4} \>\rm{n}_{\rm{e}}\>\rm{T}_{\rm{e}}^{-\frac{1}{2}}}
	\end{aligned}
\end{equation}
where F is the flux of each line and $T_{e}$ is the electron temperature. 

We measure these line ratios for each CL emitting spaxel and use \textbf{getCrossTemDen} to determine the CLR temperatures and densities in our sample. The MaNGA DAP provides flux values for the [OIII] $\lambda$4959, [OIII] $\lambda$5007, [SII] $\lambda$6717, and [SII] $\lambda$6731 lines. However, it does not provide flux values for the [OIII] $\lambda4363$ line. We thus modify our spectral fitting routine outlined in Section \ref{subsec:specfit} to measure [OIII] $\lambda4363$ in the CL emitting spaxels, if present. We note that the [OIII] $\lambda4363$ line is often blended with the H$\gamma$ line; as a result, we use a double Gaussian fit, where appropriate, to isolate [OIII] $\lambda4363$ emission. We show an example of a double Gaussian fit on the H$\gamma$ and [OIII] $\lambda4363$ lines in Appendix \ref{appendixa}.

\subsection{Galaxy Morphology}
To determine the morphological diversity of our sample and to uncover the link, if any, between CL emission and galaxy type, we use the MaNGA Morphologies from Galaxy Zoo value-added catalog, which features data from Galaxy Zoo 2 (GZ2; \citealt{2013MNRAS.435.2835W}). GZ2 is a citizen science campaign with more than 16 million visual morphological classifications for $>$ 304,000 galaxies in SDSS. 

We use the weighted vote fraction (outlined in \citealt{2013MNRAS.435.2835W}), which accounts for voter consistency, to identify CL galaxies as either elliptical (``smooth"; ``E'' in Table \ref{tab:cltable}) or spiral (``features or disks"; ``S'' in Table \ref{tab:cltable}). We also use this fraction to determine if a CL galaxy is in the process of merging (``m'' in Table \ref{tab:cltable}). We require a weighted vote fraction $\ge$ 50\% for each morphological classification. 4/10 CL galaxies in our sample do not have a GZ2 morphological classification and their morphologies are not apparent from a visual inspection (labeled `-' in Table \ref{tab:cltable}).
\subsection{AGN Bolometric Luminosity}\label{bolo}
Fueled by the accretion of gas onto supermassive black holes, AGNs emit their energy across the entire electromagnetic spectrum. The AGN bolometric luminosity can thus be a useful gauge for tracing the strength of their emission. If the bulk of CL emission is powered by AGN activity, we anticipate that the AGN bolometric luminosities of the CL galaxies will scale with their CL luminosities.

We determine the AGN bolometric luminosity for each CL-emitting spaxel using the [OIII] $\lambda$5007 flux values provided by the MaNGA DAP, and the procedure outlined in \cite{2017MNRAS.468.1433P}, which assumes [OIII] $\lambda$5007 emission comes from the AGN: 
\begin{equation}\label{eqn:00}
\begin{aligned}
\rm{log}\>\left(\frac{L_{\rm{bol}}}{erg s^{-1}}\right) = (0.5617 \pm 0.0978)\>\rm{log}\> \left(\frac{L_{[OIII]}}{erg s^{-1}}\right)\\ + (21.186 \pm 4.164)
\end{aligned}
\end{equation}
\section{Results} \label{sec:results}
In this section, we review the strength and spatial distribution of the CLs in our sample, investigate the role of SNR shocks in the production of the CLs, and present electron temperature, electron density, and AGN bolometric luminosity measurements for the CLR. We then compare our sample with the largest existing MaNGA AGN catalog to determine if CL emission is unambiguously linked to AGN activity. Finally, we cross-match our sample with three existing MaNGA BPT AGN catalogs to help uncover if CLs can help identify AGNs in large spectroscopic surveys. 

Overall, we detect 10 CL galaxies in our scan of MaNGA's MPL-8 (0.16\% of the catalog) with CL emission detected at $\ge$ 5$\sigma$ above the continuum in $>$ 10 spaxels. This detection rate suggests that we will identify $\sim$ 16 CL galaxies in the full survey of $\sim$ 10,000 galaxies. We provide the entire sample of CL galaxies in Table \ref{tab:cltable}. 


 We measure the CLR to extend between 1.3 - 23 kpc from the galactic center, with an average distance of 6.6 kpc (distances traditionally associated with the NLR).  In Appendix \ref{appendixb}, we present each CL galaxy's [OIII] $\lambda$5007 flux map to demonstrate the similar physical scales (several kpc) of the CLR and NLR. 
 
 Further, to ensure that CL emission is sufficiently resolved for each CL galaxy, we investigate the instrument PSF ($\sim$ 2.$^{\prime\prime}$5 for MaNGA). We determine that the PSF for our sample varies between 536 pc to 3.2 kpc, with an average PSF of 1.7 kpc. 6/10 CL galaxies show well resolved and continuous emission beyond the instrument PSF. The remaining four CL galaxies, J$0752$ ([NeV] $\lambda3427$), J$1535$ ([FeVII] $\lambda6086$), J$1714$ ([NeV] $\lambda3347$), and J$2051$ ([NeV] $\lambda3427$) lack continuous CL emission within a 2.$^{\prime\prime}$5 x 2.$^{\prime\prime}$5 FoV. As a result, we consider the CLRs within these galaxies to be slightly below the instrument PSF, and not spatially resolved in our analysis. The CL emission in these galaxies may indeed still be spatially resolved (e.g., CL emission may be oriented along an ionization cone); however, we reason that the CLDs we present for these galaxies are upper estimates.
\begin{table*}
	\renewcommand{\thetable}{\arabic{table}}
	\centering
	\caption{MaNGA CL Galaxies}. 
	\tabcolsep=1.0pt
	\setlength{\tabcolsep}{5pt}

	\begin{tabular}{ccccccccccccc}
		\hline
		\hline
		SDSS  & Redshift & Mor. &Detected  & L$_{\rm{Bol}}$  &   L$_{\rm{CL}}$ & CLD  & SNR & $\alpha$\\ {Name} & {(\textit{z})} & {} & {CL} &  {($10^{44}\>\rm{erg}\>\rm{s}^{-1}$)} & {($10^{40}\>\rm{erg}\>\rm{s}^{-1}$)} &  {(kpc)}  &  {Frac} & {} \\
		 {(1)} & {(2)} & {(3)} & {(4)} &  {(5)} & {(6)} &  {(7)}  &  {(8)} & {(9)} \\
		\hline
J073623.13+392617.7 & 0.12 & - & [NeV] $\lambda$3427 & 21.5$\pm$0.1 & 5.2$\pm$0.4 & 5.4$\pm$1.1 & - & -0.6$\pm$0.1 \\J075217.84+193542.2 & 0.12 & - & [NeV] $\lambda$3427 & 49.0$\pm$0.2 & 9.8$\pm$0.4 & 7.1$\pm$1.1 & 1.0 & -1.5$\pm$0.1 \\J090659.46+204810.0 & 0.11 & S(m) & [FeVII] $\lambda$3586 & 4.9$\pm$0.1 & 3.70$\pm$0.03 & 10.0$\pm$1.0 & 0.36 & -0.46$\pm$0.04 \\J134918.20+240544.9 & 0.02 & - & [FeVII] $\lambda$3760 & 18.4$\pm$0.1 & 7.20$\pm$0.01 & 7.0$\pm$0.2 & 0.38 & -0.84$\pm$0.03 \\J145420.10+470022.3 & 0.13 & E(m) & [FeVII] $\lambda$3760 & 0.75$\pm$0.09 & 2.30$\pm$0.02 & 23.0$\pm$1.1 & 0.5 & 0.2$\pm$0.1 \\J153552.40+575409.4 & 0.03 & E & [FeVII] $\lambda$6086 & 7.76$\pm$0.04 & 0.23$\pm$0.01 & 1.3$\pm$0.3 & - & -1.3$\pm$0.2 \\J161413.20+260416.3 & 0.13 & - & [NeV] $\lambda$3347 & 78.6$\pm$0.3 & 5.7$\pm$0.3 & 5.9$\pm$1.2 & - & -0.6$\pm$0.1 \\J161413.20+260416.3 & - & - & [NeV] $\lambda$3427 & 126.0$\pm$0.3 & 23.0$\pm$0.3 & 8.3$\pm$1.2 & - & -1.2$\pm$0.1 
\\J171411.63+575834.0 & 0.09& E & [NeV] $\lambda$3347 & 5.5$\pm$0.1 & 0.95$\pm$0.10 & 4.3$\pm$0.9 & - & -0.9$\pm$0.2 \\J171411.63+575834.0 &- & - & [NeV] $\lambda$3427 & 9.7$\pm$0.1 & 5.5$\pm$0.1 & 3.7$\pm$0.9 & - & -0.9$\pm$0.1 \\J205141.54+005135.4 & 0.11 & S & [NeV] $\lambda$3427 & 4.3$\pm$0.1 & 0.75$\pm$0.03& 2.5$\pm$1.0 & - & -0.6$\pm$0.1 \\J211646.34+110237.4 & 0.08 & S & [NeV] $\lambda$3427 & 41.4$\pm$0.1 & 3.1$\pm$0.1 & 4.3$\pm$0.8 & 1.0 & -1.2$\pm$0.1 \\

		\hline 
		\multicolumn{9}{p{\textwidth}}
	{Note: Columns are (1) CL galaxy SDSS name, (2) redshift from the NASA Sloan Atlas catalogs using single-fiber measurements \citep{2011AJ....142...31B}, (3) Galaxy Zoo 2 morphology (if available), (4) detected CL in the galaxy, (5) AGN bolometric luminosity estimated from [OIII] measurements (\citealt{2017MNRAS.468.1433P}), (6) CL luminosity measured in this work, (7) galactocentric distance of the most extended CL-emitting spaxel, (8) fraction of SNRs in the CLR (if available), and (9) slope of the CL intensity profile.}
	\end{tabular}
	\label{tab:cltable}
\end{table*}  
\subsection{The Strength and Distribution of the CLs}\label{CL Flux Maps}
Extended emission line regions (1 kpc to 100 kpc from the galactic center) surrounding AGNs are expected to feature a variety of ionization mechanisms, which include photoionization and shocks (e.g., \citealt{2002RMxAC..13..213T}). To determine the role that these ionization sources play in the production of CLs, we explore the strength and distribution of the CLs.


%
\subsubsection{[NeV] $\lambda3347$, $\lambda3427$ }\label{nevsection}
The [NeV] $\lambda3347$ line is produced from the same excitation level as  [NeV] $\lambda3427$, but its relative flux \big(100 $\times$ CL Flux/Flux (Ly$\alpha$); based on a composite quasar spectrum from \citealt{2001AJ....122..549V}\big) is 0.118 $\pm$ 0.008, approximately one-third of the [NeV] $\lambda3427$ line (0.405 $\pm$ 0.012; \citealt{2001AJ....122..549V}). As a result, we expect to detect the stronger [NeV] $\lambda3427$ line in a higher fraction of MaNGA galaxies, which we do. 
	
\begin{figure*}
	\includegraphics[width=13.5cm,height=4.5cm]{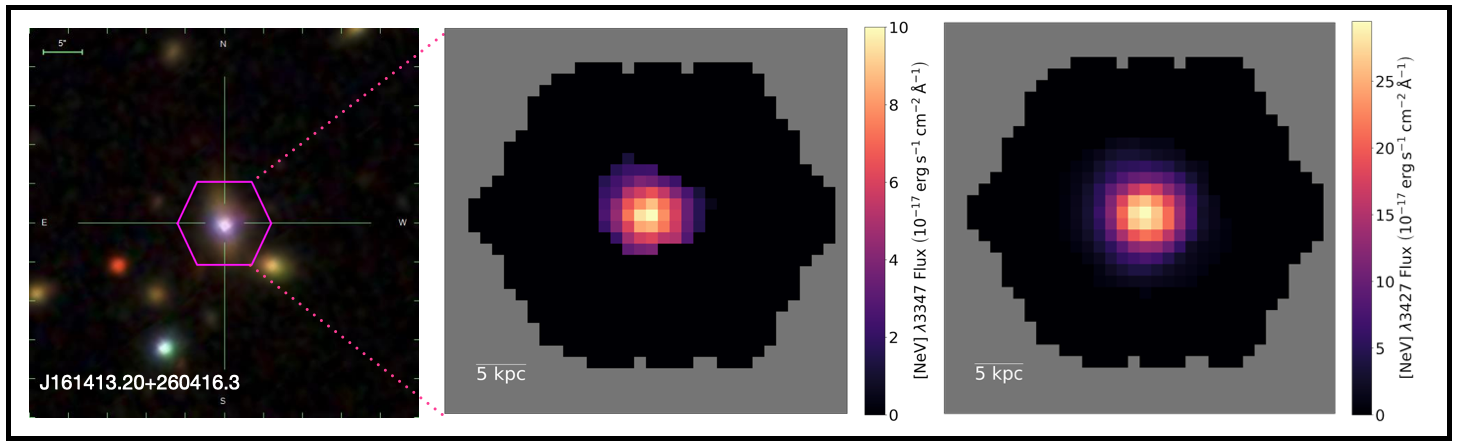}
	\includegraphics[width=13.5cm,height=4.5cm]{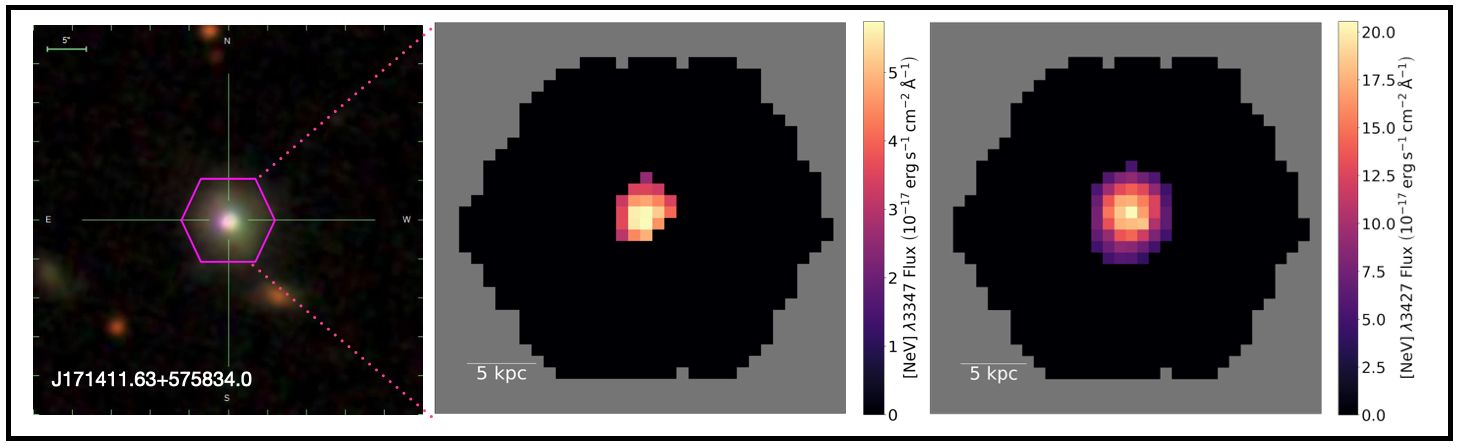}
	\centering
	\caption{From top to bottom: SDSS optical image and CL flux maps for J$1614$ ([NeV] $\lambda 3347$, $\lambda 3427$) and J$1714$ ([NeV] $\lambda 3347$, $\lambda 3427$).}
	
	\label{nevmaps}
\end{figure*}

We  measure [NeV] $\lambda3427$ emission in six galaxies from  MPL-8: J$1614$,  J$1714$,  J$0736$, J$0752$,  J$2051$, and J$2116$. In two of these galaxies, J$1614$ and J$1714$, we also measure [NeV] $\lambda3347$ emission. We present [NeV] $\lambda 3347$, $\lambda 3427$ flux maps for J$1614$ (\textit{z} = 0.13;  no GZ2 classification) and J$1714$ (\textit{z} = 0.09; elliptical) in Figure \ref{nevmaps}. For J$1614$, we identify extended [NeV] $\lambda 3347$ emission out to 5.9  $\pm$ 2.9 kpc from the galactic center; 8.3 $\pm$ 3.9 kpc for [NeV] $\lambda 3427$. For J$1714$, we measure [NeV] $\lambda 3347$ emission out to 4.3 $\pm$ 2.2 kpc from the galactic center; 3.7 $\pm$ 2.2 kpc for [NeV] $\lambda 3427$. We show the [NeV] $\lambda 3427$ flux maps for J$0736$ (\textit{z} = 0.12; no GZ2 classification; CLD = 5.4 $\pm$ 2.7 kpc), J$0752$ (\textit{z} = 0.12; no GZ2 classification; CLD = 7.1 $\pm$ 2.7),  J$2051$ (\textit{z} = 0.11; spiral; CLD = 2.5 $\pm$ 2.5 kpc), and J$2116$ (\textit{z} = 0.08; spiral; CLD = 4.3 $\pm$ 1.9 kpc) in Figure  \ref{fevmaps}. In each galaxy, [NeV] $\lambda 3347$, $\lambda 3427$ emission is predominantly concentrated within the galactic interior (within $ 2.^{\prime\prime}$5 throughout the paper), suggesting that a central source governs CL production in these galaxies.
We cross-match these galaxies with the largest existing MaNGA AGN catalog (``MaNGA AGN catalog''; \citealt{2020arXiv200811210C}; AGN classifications determined using \textit{WISE} mid-infrared color cuts, \textit{Swift}/BAT hard X-ray observations, NVSS/ FIRST 1.4 GHz radio observations, and SDSS broad emission lines; Section \ref{tab:mangaagn}), and determine that all [NeV] $\lambda 3347$, $\lambda 3427$ galaxies are classified as AGNs. 

%
\newpage
\subsubsection{[FeVII] $\lambda 3586, \lambda 3760, \lambda 6086$ }
\label{sec:fev}
\cite{2001AJ....122..549V} analyzed a homogeneous data set of over 2,200 quasar spectra from SDSS and created a variety of composite spectra from this sample. The authors determined that the relative fluxes for [FeVII] $\lambda3760, \lambda3586, \lambda6086$ are 0.078 $\pm$ 0.007, 0.100 $\pm$ 0.011, and 0.113 $\pm$ 0.016, respectively ($\sigma = 0.018$), which feature significantly less scatter than the [NeV] $\lambda$3347, $\lambda3427$ lines ($\sigma = 0.20$). As a result, we expect to find a similar fraction of [FeVII] galaxies that emit each line, which we do. 

\begin{figure*}
	\includegraphics[width=\textwidth,height=4.5cm]{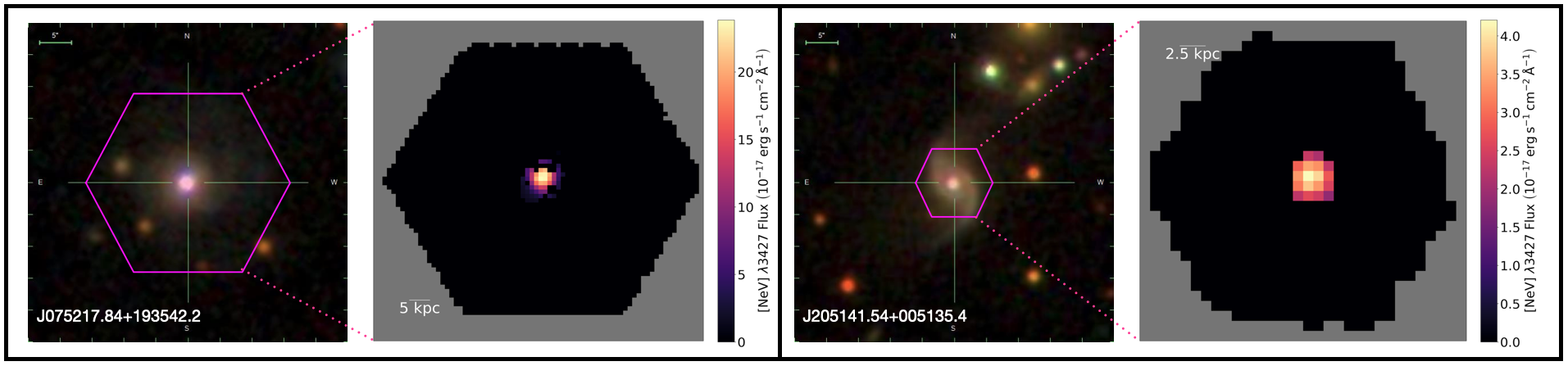}
	\includegraphics[width=\textwidth,height=4.5cm]{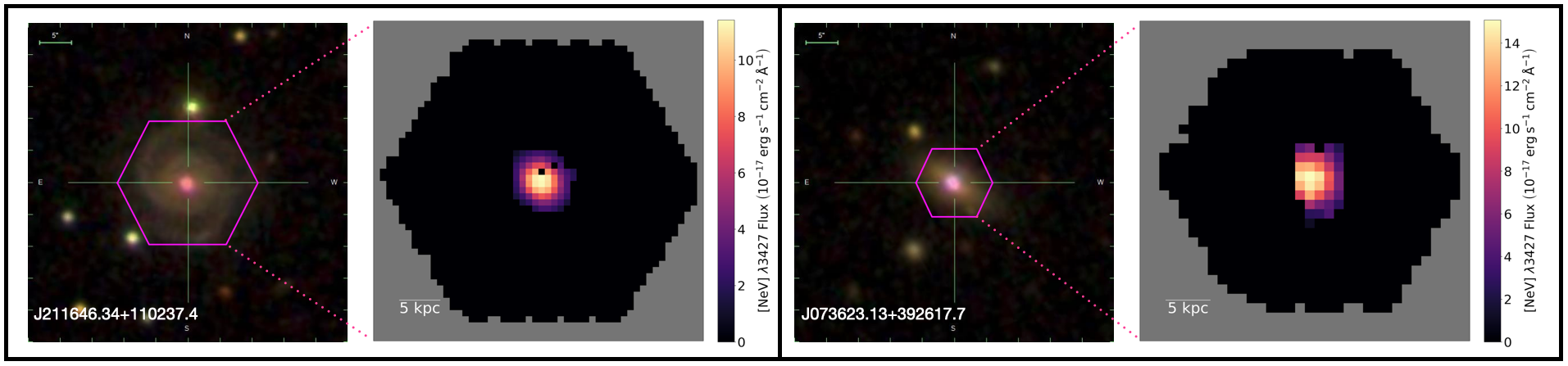}\\
	\includegraphics[width=\textwidth,height=4.5cm]{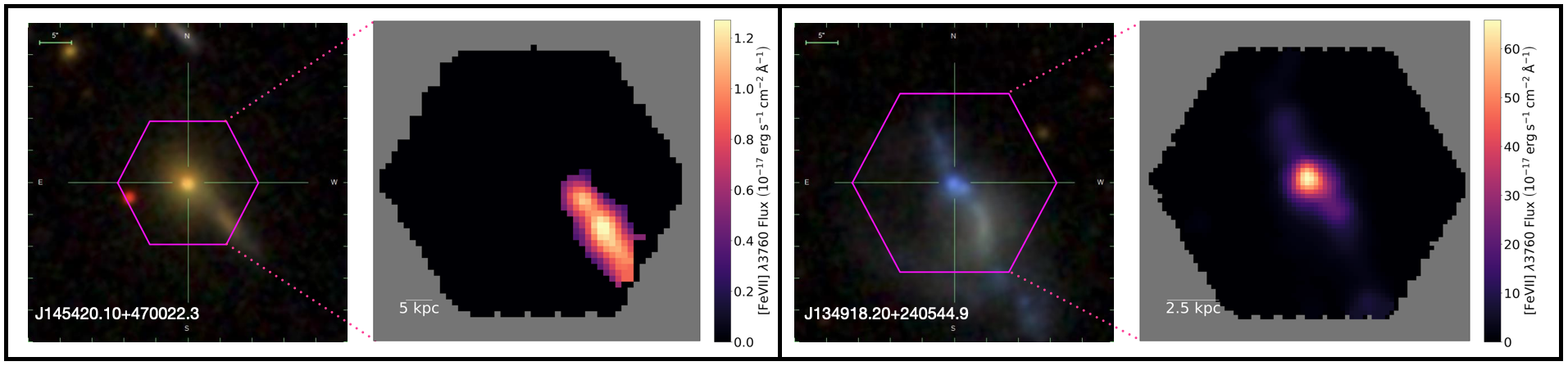}\\
	\includegraphics[width=\textwidth,height=4.5cm]{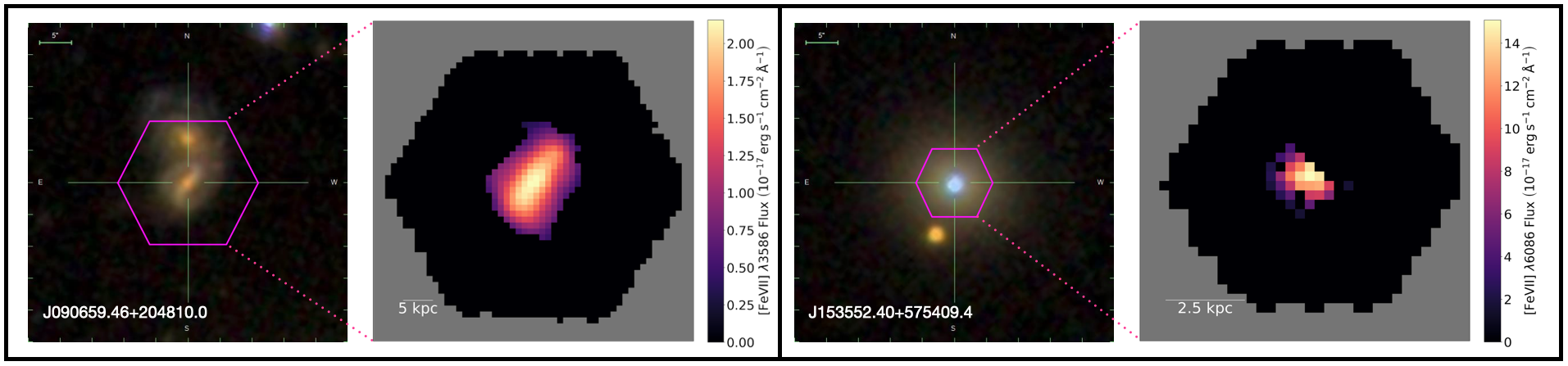}
	\caption{From left to right and top to bottom: the SDSS optical image and CL flux maps for J$0752$, J$2051$, J$2116$, J$0736$, J$1454$, J$1349$, J$0906$, and J$1535$. The CLDs for the CL mergers are the most extended in our sample (10.0 $\pm$ 2.5 kpc for J$0906$ and 23.0 $\pm$ 2.9 kpc for J$1454$). Merger-induced shocks are likely to be the primary ionization mechanism for these galaxies. The CL galaxies that host AGNs tend to have more compact CL emission that is predominantly within the galactic interior, which suggests AGN photoionization is the primary CL ionization mechanism for these galaxies.}
	\label{fevmaps}
\end{figure*}

We measure [FeVII] emission in four MPL-8 galaxies: J$1454$ ([FeVII] $\lambda3760$; \textit{z} = 0.13; elliptical; merger), J$1349$ ([FeVII] $\lambda3760$; \textit{z} = 0.02; no GZ2 classification),  
J$0906$ ([FeVII] $\lambda3586$; \textit{z} = 0.11; spiral; merger), and J$1535$ ([FeVII] $\lambda6086$;  \textit{z} = 0.03; spiral). We present the flux distributions for these galaxies in Figure \ref{fevmaps}. 

Previous studies (e.g., \citealt{1981ApJ...246..696O, 1984ApJ...286..171D, 1986ApJ...301..727D, 1984ApJ...285..458F}) posit that the CLR lies within a transition zone between the BLR ($\sim$ 0.1 kpc) and the NLR (several kpc). Here, we measure [FeVII] $\lambda 3760$ flux in J$1454$ out to 23.0 $\pm$ 2.9 kpc (4.4$\sigma$ above the mean of the CLD distribution; the largest in our sample). We also find that the bulk of the CLR in this galaxy is not spatially coincident with the galactic center. We determine that this galaxy is classified as a merger in GZ2 and we detect the companion southwest of the galactic center, well-aligned with the [FeVII] $\lambda 3760$ emission. During the merging process, companion galaxies can be separated by tens of kpcs (e.g., \citealt{2010ApJ...724..267F}), which we reason likely accounts for the far-reaching CL emission in J$1454$ (see Section \ref{CLmerg}).

We measure [FeVII] $\lambda3760$ emission in J$1349$ out to 7.0 $\pm$ 0.6 kpc from the galactic center. The CL emission in this galaxy is most abundant within the galactic interior, but also extends throughout the southwest and northeast regions. No GZ2 classification is available for this galaxy, but its strong CL emission (1.2$\sigma$ above the mean of the CL luminosity distribution; Table \ref{tab:cltable}) and its extended morphology make it a galaxy of peculiar interest in our analysis. We further analyze this galaxy in Section \ref{sec:bolo} and determine it to be an AGN candidate. 

For J$0906$, we identify substantial [FeVII] $\lambda 3586$ emission out to 10.0 $\pm$ 2.5 kpc from the galactic center (1.9$\sigma$ above the mean of the CLD distribution; the second largest in our sample). We determine that this galaxy is a merger based on the GZ2 classification and identify elongated CL emission along an apparent central bar. We detect the companion in the northern region of this galaxy, partially within the MaNGA FoV.

We scan the MaNGA AGN catalog and find no AGN classification for J$1454$, J$1349$, or J$0906$ (two mergers; one unclassified morphology). We consider the likelihood that these galaxies are AGN candidates, to help determine if AGN processes are producing the CLs in these galaxies, in Section \ref{sec:bolo}. 

Moreover, prior studies show that gas inflows produced by mergers can generate widespread shocks that ionize gas on kpc and sub-kpc scales (e.g., \citealt{2015MNRAS.448.2301M}). As such, we also consider the influence of a companion galaxy to be a potential source of the far-reaching CL emission in the merging galaxies (Section \ref{CLmerg}). 
%

For J$1535$ (a confirmed AGN in the MANGA AGN catalog), we detect [FeVII] $\lambda 6086$ emission out to 1.3 $\pm$ 0.76 kpc. The CL distribution in J$1535$ strongly resembles the nuclear-bound [NeV] $\lambda3347,3427$ distributions (Section \ref{nevsection}). If CLs are produced predominately by AGN photoionization, then we expect to find the bulk of CLs within the galactic interior, which is the case for this galaxy.
\subsubsection {CL AGNs and CL Mergers}
\label{CLmerg}
We find that the CL galaxies in our sample are either hosting an AGN (7; ``CL AGNs"), undergoing a merger (2; ``CL mergers"), or are unclassified (1; an AGN candidate;  Section \ref{sec:bolo}). The strength and distribution of the CL AGNs are consistent with AGN photoionization playing the most direct role in the production of the CLs, with merger-induced shocks being a strong candidate for ionization, too. The CL AGN detection rate (70\% - 80\%) suggests that CLs are useful for identifying AGN in large spectroscopic surveys. 


Additionally, we measure the CLDs of the CL mergers to be the most extended in our sample (10.0 $\pm$ 2.5 kpc for J$0906$ and 23.0 $\pm$ 2.9 kpc for J$1454$). The interactions between companion galaxies, which occur on the scale of tens of kpcs, can create tidal forces that drive gas towards the galactic centers, which can lead to shock excitation (e.g., \citealt{2010ApJ...724..267F}). These shocks are likely to impact both galaxies and account for the extended CL emission we see in the CL mergers. Moreover, the CL mergers may be relevant for AGN studies because infalling gas can fuel AGN activity, which can produce large-scale outflows and additional shocks that influence the host galaxy's evolution (e.g., \citealt{2015ApJS..221...28R}).

\subsection{The Role of Stellar Shocks in the CLR}
\label{sec:shocks}
Shock ionization can result from a variety of phenomena, including, but not limited to, AGN activity (e.g., jets), gas inflows produced by mergers, and stellar winds generated by SNRs (e.g., \citealt{2019ApJ...880...16K}). 
We address the role of SNR shock ionization in the CLR, specifically, by investigating the populations of SNRs in the CLR. To execute this analysis, we adopt the approach outlined in Section \ref{shockssec}.

\begin{figure}[ht!]
	\centering
	{{\includegraphics[width=8cm]{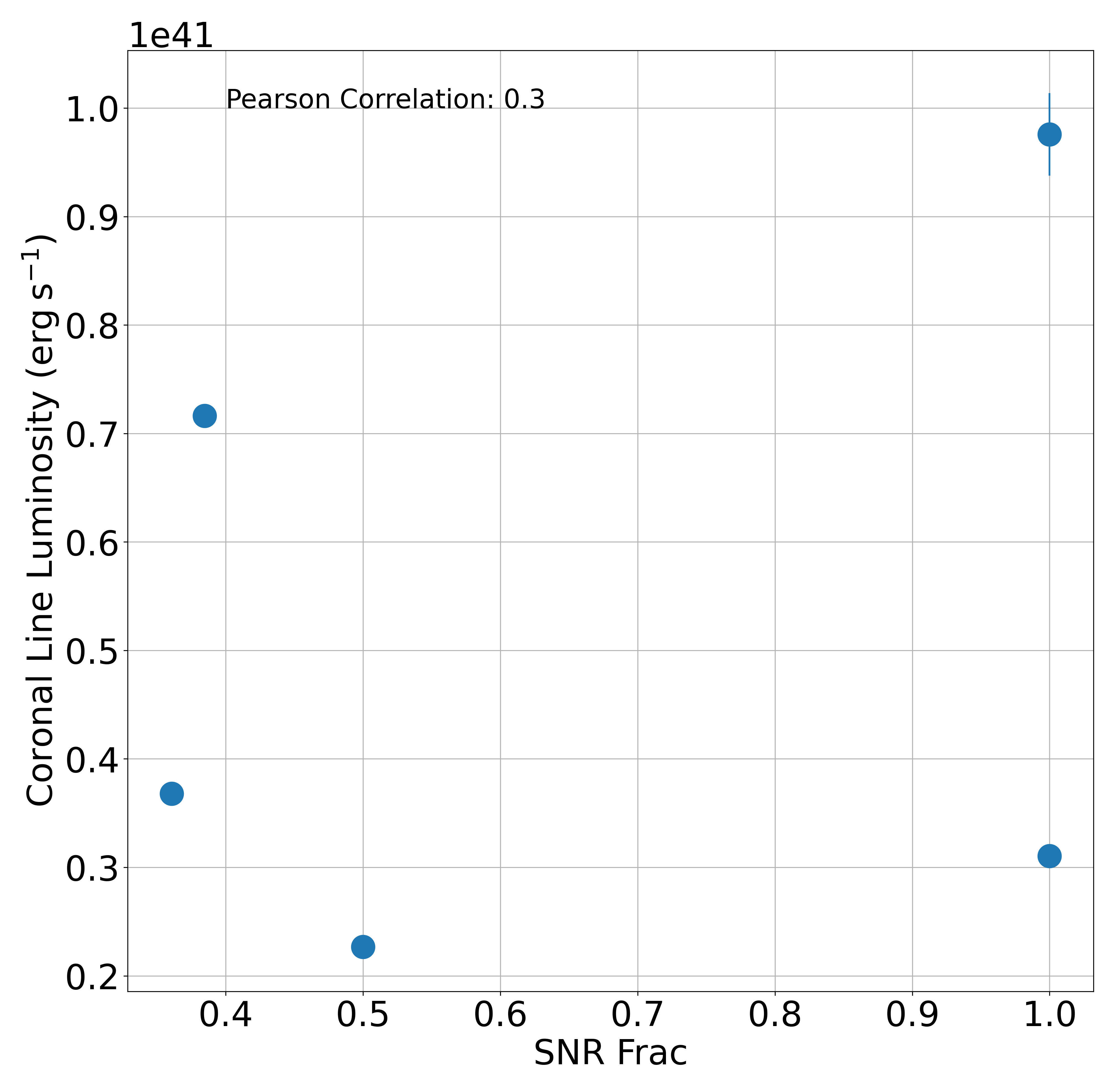} }}%
	\qquad
	\caption{Total CL luminosity for each CL galaxy (with SNR emission in at least one CL spaxel) vs. the fraction of SNRs found in their respective CLRs. We measure a Pearson correlation value of 0.3, which suggests that, on average, higher fractions of SNRs do not produce more abundant CL emission. Due to the few data points and the lack of a clear visual trend, we consider these results preliminary.  For some data points, CL luminosity uncertainties are too small to be visualized on the plot.}%
	\label{fig:snr_cl}
\end{figure}

Further, to better decipher the role of SNRs in the CLR, we plot the CL luminosity for each galaxy in our sample against the fraction of SNRs in their CLRs (Figure \ref{fig:snr_cl}). We utilize the Pearson correlation coefficient to quantify our results. The coefficient ranges from -1 to +1, where 0 implies no correlation, -1 is indicative of a negative correlation, and +1 signifies a positive correlation. If CL luminosity and the fraction of SNRs are positively correlated, then SNRs likely play a critical role in the production of CLs.

We compute a Pearson correlation value of 0.3, implying a weak positive correlation between the fraction of SNRs in the CLR and the strength of the CLs. This suggests that a higher fraction of SNR shocks, on average, do not increase the strength of CL emission. We reason that AGN photoionization and merger-induced shocks are the dominant ionization mechanisms producing the CLs, even for the CL galaxies that feature a 100\% fraction of SNRs in their CL-emitting spaxels (Section \ref{sec:bolo}). However, we do acknowledge the sparse amount of data in Figure \ref{fig:snr_cl} and a lack of a clear visual trend. While we expect photoionization and merger-induced shocks to have the most impact on the strength of the CLs, further analysis is required to definitively rule out the influence of SNRs in the CLR. 

We also inspect the ionizing radiation field of the CL galaxies to determine the ionization mechanism(s) producing the CLR. We consider the r$^{-2}$ ionizing radiation dilution characteristic of pure AGN photoionization (e.g., \citealt{2002RMxAC..13..213T, 2012ApJ...747...61Y}; Section \ref{sec:power-law}), and compare the power law indexes of the CL luminosities to $\alpha$ = -2 (Table \ref{tab:cltable}). We measure $\alpha$ to range between -1.8 $\pm$ 0.3 to -0.6 $\pm$ 0.1 for the CL AGNs (7) and -0.9 $\pm$ 0.1 to 0.2 $\pm$ 0.1 for the remaining sample (3). For both samples, the decay of ionizing radiation with increasing galactocentric distances is shallower than the profile expected for pure AGN photoionization. As a result, we reason that the CLs, even for the CL AGNs, likely feature a blend of ionization mechanisms that include, but are not limited to, SNRs and merger-induced shocks, which account for the extended nature of the CLR (Section \ref{sec:results}). Further, we acknowledge that fluctuations in the density profile, as a function of galactocentric distance (Section \ref{Temp}; e.g., \citealt{2021ApJ...910..139R}), may  influence the evolution of the ionizing radiation field, and that merger-driven shocks, AGN photoionization, and SNR shocks may not be the only CL ionization mechanisms; stochastic accretion, AGN light echoes, and AGN outflows are also possible mechanisms that can produce CL emission (e.g., \citealt{2011ApJ...739...69M,2016arXiv160404515W}). 

\newpage
\subsection{Electron Temperatures and Electron Number Densities of the CLs}\label{Temp}
Previous studies (e.g., \citealt{2010MNRAS.405.1315M}) found that temperatures in the CLR range from 10,000 K - 20,000 K, consistent with pure AGN photoionization. However, beyond the $\sim$ 20,000 K threshold, up to 10$^{6}$ K, outflowing jets can collisionally ionize and heat gas clouds (via stellar and AGN shocks, for example; see \citealt{2002A&A...383...46M}; Section \ref{tempintro}). For these cases, kinetic energy must be supplied by an additional source (e.g., \citealt{1984QJRAS..25....1O}). Moreover, intermediate electron densities $n_{e}$ $\approx$ 10$^{2}$ - 10$^{4}$ cm$^{-3}$ are typical of the NLR; $n_{e} >$ 10$^{9}$ cm$^{-3}$ for the BLR. 

\begin{figure*}
	\includegraphics[width=\textwidth,height=4.2cm]{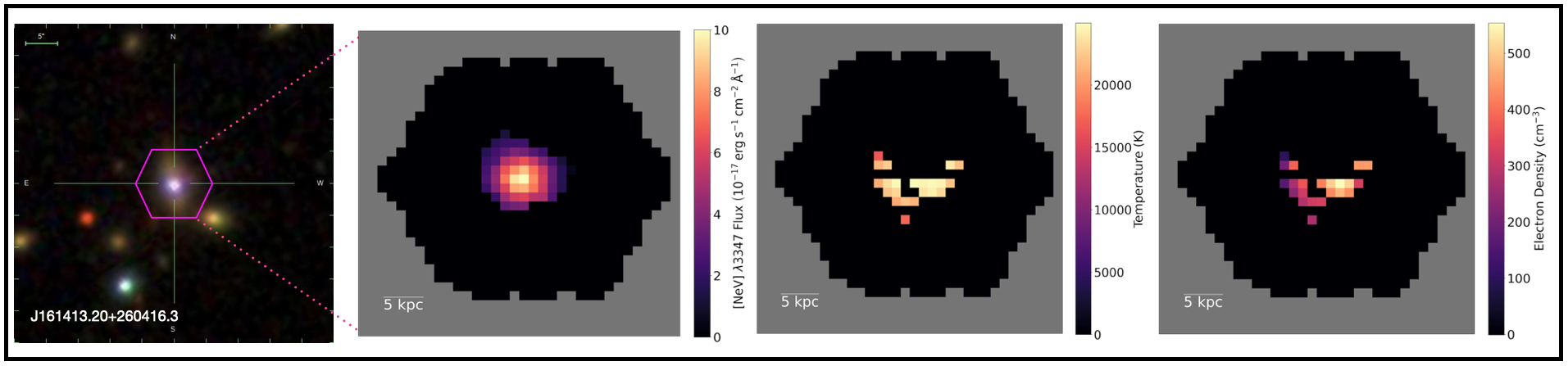}
	\includegraphics[width=\textwidth,height=4.2cm]{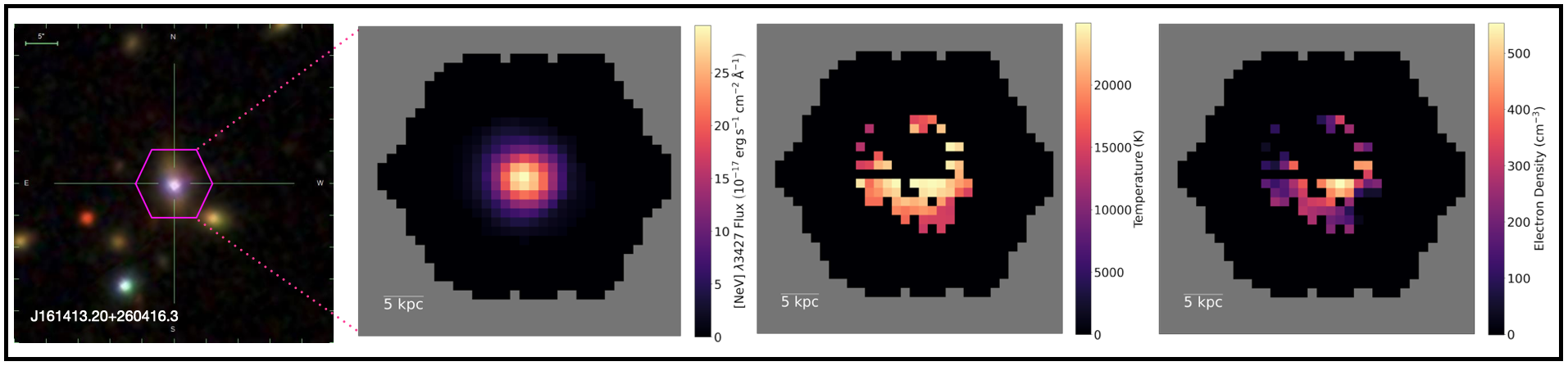}\\
	\includegraphics[width=\textwidth,height=4.2cm]{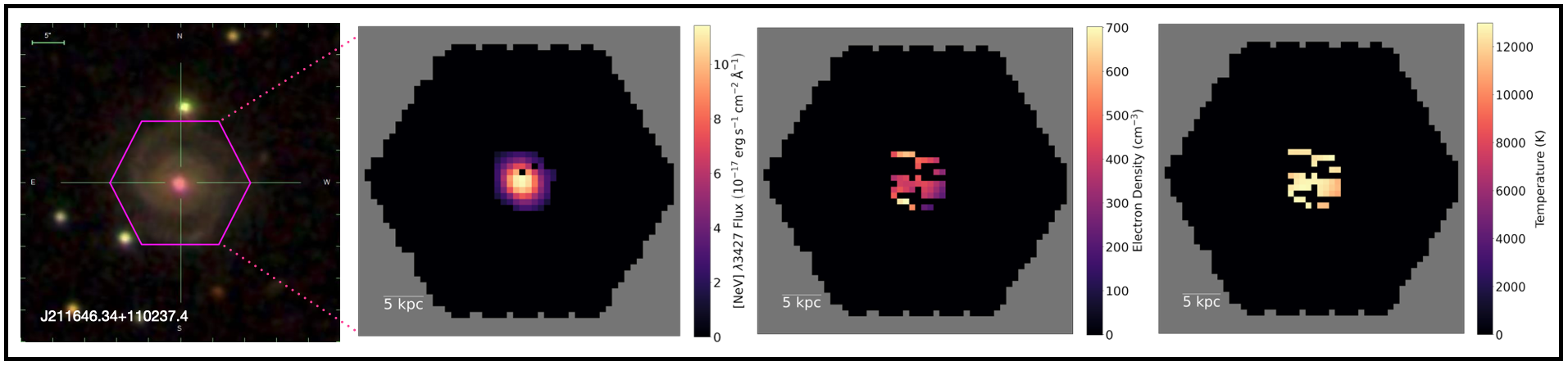}\\
	\	\includegraphics[width=\textwidth,height=4.2cm]{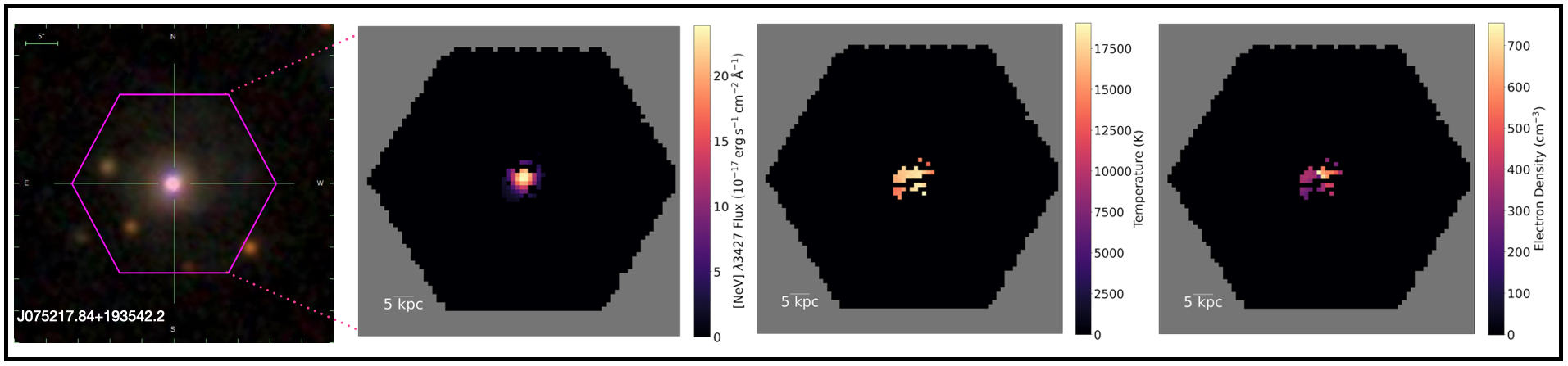}
	
	\caption{Left to right and top to bottom: SDSS optical image, and CL flux, temperature, and density maps for J$1614$ ([NeV] $\lambda 3347$), J$1614$ ([NeV] $\lambda 3427$), J$2116$ ([NeV] $\lambda 3427$), and J$0752$ ([NeV] $\lambda 3427$).}
	\label{temdenmaps}
\end{figure*}

As a result, we measure the temperature and density profiles for the galaxies in our sample to better understand the nature of the CLR (Figure \ref{temdenmaps}). For the density profiles, we acknowledge that recent studies (e.g.,  \citealt{2019MNRAS.486.4290B, 2020MNRAS.498.4150D}) report that electron number densities derived from the [SII] $\lambda\lambda$6717,6731 doublet may be underestimated. Thus, we consider our electron density measurements to be lower estimates.

For the temperature and density measurements, we require flux values for the [OIII] $\lambda$4363 line, but we only detect this line in 4/10 CL galaxies, which all host AGNs (Table \ref{tab:tempden}). We also consider the fraction of spaxels in each galaxy's CLR that we can measure temperatures and densities for (the ``Measurement Fraction" in Table \ref{tab:tempden}). We find that the average CLR temperatures vary between 12,331 - 22,530 K, and that the average temperatures for 2/4 CL galaxies are consistent with pure photoionization. However, 2/4 feature temperatures slightly above the $\sim$ 20,000 K threshold (for J$1535$, we determine a temperature of 21,088 K in the single spaxel we could measure). We reason that for both of the CL AGNs with average temperatures above this limit, shock-induced compression and heating contribute to the production of the CLs (see \citealt{2021MNRAS.501L..54R} for a similar analysis of shocked emission in AGNs). 

Additionally, we determine that the average CLR density varies between 244 - 586 cm$^{-3}$, typical of the NLR. This result is consistent with the CLDs extending well into the NLR (Section \ref{CL Flux Maps}). 
\begin{table*}
	\renewcommand{\thetable}{\arabic{table}}
	\centering
	\caption{Average CLR Electron Temperatures and Number Densities}
	\begin{tabular}{ccccc}
		\hline
		\hline
		SDSS Name & CL & Electron Temperature  & Electron Density & Measurement Fraction\\ {} & {} & {(K)} & {(cm$^{-3}$)} & {} \\
		
		{(1)} & {(2)} & {(3)} & {(4)} &  {(5)} \\
		\hline
		J075217.84+193542.2 &[NeV] $\lambda$3427 & 16,680 & 400 &56\% \\
		J153552.40+575409.4& [FeVII] $\lambda$6086  &   21,088 & 586 & 3\%\\
		J161413.20+260416.3&   [NeV] $\lambda$3347 &22,530  & 355 & 36\% \\
		& [NeV] $\lambda$3427 &   19,533  & 244 & 38\%\\
		J211646.34+110237.4&  [NeV] $\lambda$3427 &12,331  & 414 & 51\%\\
		\hline
		\multicolumn{5}{p{\textwidth}}
		{Note: Columns are (1) CL galaxy SDSS name, (2) detected CL, (3) average CLR electron temperature, (4) average CLR electron temperature, and (5) the fraction of spaxels in each galaxy's CLR that we can measure temperatures and densities for.}
		\label{tab:tempden}
	\end{tabular}
\end{table*}  

\subsection{AGN Bolometric Luminosity}\label{sec:bolo}
\begin{figure}[t]
	\includegraphics[height=7.5cm]{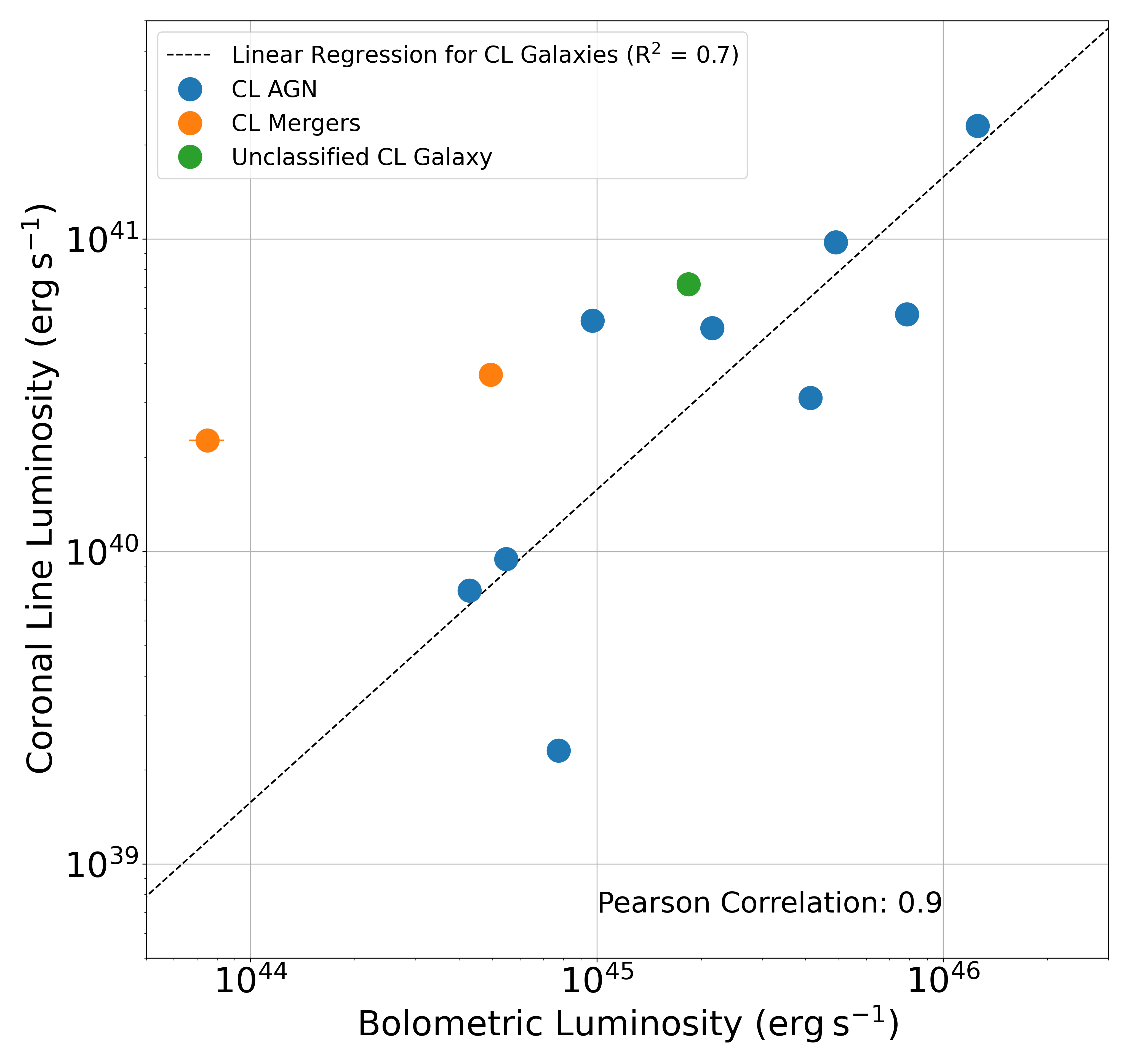}
	\caption{Total CL luminosity plotted against bolometric luminosity for all CL galaxies. The black line in the linear regression fit to the data. The blue data points are the CL AGNs, the orange are the CL mergers, and the green is the unclassified CL galaxy J1349.  For some data points, CL and bolometric luminosity uncertainties are too low to be visualized on the plot.}
	\label{fig:bolo}
\end{figure}

AGN bolometric luminosity measures the total luminosity emitted by an AGN across all wavelengths. It effectively traces the accretion efficiency, which is the fraction of accreted mass on the SMBH that is radiated, and offers direct insight into AGN power (e.g.,  \citealt{2009MNRAS.396.1217R}). If CLs provide definitive evidence for AGN activity, then the CL luminosity of each CL-emitting spaxel should broadly scale with its bolometric luminosity. We explore this relation using the method outlined in Section \ref{bolo}, and present our results in Figure \ref{fig:bolo}. For galaxies with multiple CL detections, we plot each CL independently. We identify a strong positive correlation between CL and bolometric luminosities for our full sample (Pearson value of 0.9). In general, the strength of the CLR (measured by CL emission) scales with bolometric luminosity. 

We then perform a linear regression on the full sample of CL galaxies. We use the R$^{2}$ statistic to quantify our results. This is a statistical measure that uses the sample's variance to determine how close the data are to the fitted regression line. It varies between 0 and 1, where 0 implies high variability of the data (i.e. a poor model fit) and 1 suggests low variability (i.e. a good fit). We measure R$^{2}$ for our sample to be 0.7, which suggests that CL luminosity is dependent on AGN bolometric luminosity, for the majority of our sample. Additionally, we measure the residuals to determine how well the data fit the regression line. We calculate the quantity R$_{f}$ as the ratio of the residuals to the predicted regression values for each data point (Table \ref{tab:residual}). This dimensionless quantity measures the deviation of the data points relative to their predicted model values. 

We first analyze the CL AGNs and determine that they fit the regression well, as we expect, with R$_{f}$ values $<$ 3. We then target the CL galaxies that have not been classified as an AGN in the MaNGA AGN catalog, which are either undergoing a merger (J$1454$ and J$0906$) or are unclassified (J$1349$). We do so to identify if these are outliers in our sample of CL galaxies, and to determine if they are potential AGN candidates. J$1454$ and J$0906$ are marked by the orange data points in Figure \ref{fig:bolo}; J$1349$ in green. We measure R$_{f}$ values for J$1454$ and J$0906$ to be 8.1$\sigma$ and 1.7$\sigma$ above the mean of the $R_{f}$ distribution, respectively - the two highest in our sample. This suggests that J$1454$ is not a strong AGN candidate, and that J$0906$ requires further analysis to determine if it hosts an AGN.  On the other hand, J$1349$ ($R_{f} = 1.5$) is within 1$\sigma$ of the mean of the distribution. We consider this galaxy to be an AGN candidate due to its proximity to the regression line. 

We deduce that CLs are strong tracers of AGNs, but perhaps not perfect since we also find that galaxy mergers (which may not host an AGN) can also feature a CLR.
%


\subsection{MaNGA AGN Catalog Comparison}\label{sec:catalog}
Several authors suggest that CLs are common features in the spectra of AGNs and provide unambiguous signatures for AGN activity due to their high production energies (e.g., \citealt{1989ApJ...343..678K, 2002MNRAS.329..309P, 2010MNRAS.405.1315M, 2011MNRAS.414..218L, 2011ApJ...739...69M}). We use a catalog of confirmed AGNs in MaNGA to test the robustness of CLs as AGN identifiers (Table \ref{tab:mangaagn}).

 \cite{2020arXiv200811210C} assembled the largest existing catalog of AGNs in MaNGA to date. The authors used \textit{WISE} mid-infrared color cuts, \textit{Swift}/BAT hard X-ray observations, NVSS/ FIRST 1.4 GHz radio observations, and SDSS broad emission lines to create their sample. In total, they reported 406 unique AGNs in MPL-8. 

Mid-infrared emission, generated by the heated dust which encompasses an AGN, is a reliable probe for obscured and unobscured AGN activity. As a result, they used observations from the \textit{Wide-Field Infrared Survey Explorer} (\textit{WISE}; \citealt{2010AJ....140.1868W}) to help identify AGNs. They considered the four bands observed with \textit{WISE} (3.4 $\mu$m (W1), 4.6 $\mu$m (W2), 12 $\mu$m (W3), and 22 $\mu$m (W4)) and apply a 75\% reliability criteria of W1 - W2 $>$ 0.486 exp\{0.092(W2 - 13.07)$^{2}$\} and W2 $>$ 13.07, or W1 - W2 $>$ 0.486 and W2 $\le$ 13.07  (\citealt{2018ApJS..234...23A}) for their analysis. They identified 67 WISE AGNs in MaNGA.

To trace the hot X-ray corona around an AGN, the authors used the catalog assembled by \cite{2018ApJS..235....4O}, which consists of $\sim$ 1,000 AGNs observed by the \textit{Swift} Observatory's Burst Alert Telescope (BAT) in the ultra hard X-ray (14 - 195 keV). The authors found 17 AGNs from this catalog in MaNGA.

Further, radio studies are unique in their ability to detect strong emission emanating from AGN radio jets. \cite{2012MNRAS.421.1569B} used observations from the 1.4 GHz NRAO Very Large Array Sky Survey (NVSS; \citealt{1998AJ....115.1693C}) and the Faint Images of the Radio Sky at Twenty Centimeters (FIRST; \citealp{1995ApJ...450..559B}) to detect AGNs in the SDSS's seventh data release (DR7). They differentiated AGN activity from star formation emission using the correlation between the 4000 \AA $\>$ break strength and radio luminosity per stellar mass, emission line diagnostics, and the relation between H$\alpha$ and radio luminosity (\citealt{1995ApJ...450..559B}). \cite{2020arXiv200811210C} found 325 radio AGNs from this catalog in MaNGA. 

Broad Balmer emission lines (FWHM $>$ 1,000 km s$^{-1}$) also serve as useful signatures for AGN activity. The high velocity clouds which produce these lines provide clear evidence of high density gas in close proximity to the SMBH. \cite{2015ApJS..219....1O} assembled a catalog of nearby (\textit{z} $\le$ 2) Type I AGNs in SDSS's seventh data release using the broad H$\alpha$ emission line, and \cite{2020arXiv200811210C} identified 55 broad line AGNs from this catalog in MaNGA. 

We cross match our sample with the 406 unique MaNGA AGNs reported by \cite{2020arXiv200811210C} and find 7 CL galaxies in it (70\% of our sample). We consider CLs to be a strong tracer of AGN activity since the majority of our CL galaxies are confirmed to host an AGN. Additionally, the remaining 30\% of our sample are galaxies of interest. J$1349$ is an  AGN candidate (Section \ref{sec:bolo}), and  J$1454$ and J$0906$ are undergoing mergers, which can induce strong gas inflows that can eventually trigger the activation of AGNs and fuel large-scale AGN outflows  (e.g., \citealt{2015ApJS..221...28R}).
\subsection {BPT AGN Catalog Comparison}
 Baldwin-Phillips-Terlevich optical emission line diagnostic diagrams (BPT diagrams; \citealt{1981PASP...93....5B, 2001ApJ...556..121K, 2006MNRAS.372..961K, 2003MNRAS.346.1055K}) incorporate line ratios between high and low ionization species (e.g., [OIII] $\lambda5007/H\beta$ vs. [NII] $\lambda6584$/H$\alpha$ or [OIII] $\lambda5007/H\beta$ vs. [SII] ($\lambda6717$ + $\lambda$6731)/H$\alpha$) to distinguish gas ionization sources as star forming, Seyfert (AGN), low-ionization nuclear emission-line region (LINER), or a composite of multiple ionization sources. They serve as the traditional AGN selection tool for most spectroscopic surveys.
 
  \cite{2017MNRAS.472.4382R}, \cite{ 2018RMxAA..54..217S}, and  \cite{2018MNRAS.474.1499W} used these diagrams to identify AGN candidates in MaNGA's MPL-5, which contains data cubes for 2,727 unique galaxies. We scan these catalogs to identify the fraction of CL galaxies found within them. We do so to determine the strength of the BPT diagrams as an AGN selection tool, and to assess if CLs can help provide accurate AGN identifications in MaNGA (and similar spectroscopic surveys). We acknowledge that each BPT catalog may not have exhaustively scanned MPL-5 (e.g., spectroscopic data may not have been available for all galaxies during each analysis; \citealt{2017MNRAS.472.4382R}). As a result, we consider our findings preliminary. 
  
A primary drawback of BPT diagrams is their inability to distinguish low ionization AGNs from post-AGB stars. \cite{2017MNRAS.472.4382R} used the equivalent width of H$\alpha$ (EW H$\alpha$) to mitigate this issue. \cite{2010MNRAS.403.1036C} first determined that galaxies with EW H$\alpha$ $<3$ \AA $\>$ are predominantly ionized by post-AGB stars, rather than AGNs. As a result, \cite{2017MNRAS.472.4382R} used this threshold to analyze single-fiber (within the central 3$^{\prime\prime}$)  MPL-5 galaxy observations. They identified 62 AGNs above this limit, which were also reported as Seyfert or LINER (which have been shown to strongly link with AGN activity; e.g., \citealt{2008ARA&A..46..475H}) on the BPT diagrams.  We scan this catalog for the CL AGNs that are in MPL-5 (J$1714$; J$0736$; J$2116$; J$1535$) and determine that 0/4 are present.
\begin{table}[t]
	\renewcommand{\thetable}{\arabic{table}}
	\centering
	\caption{R$_{f}$ Values for the CL Luminosity vs. Bolometric Luminosity Relationship}
	\begin{tabular}{ccccc}
		\hline
		\hline
		SDSS Name & CL & R$_{f}$ \\ {} & {} & {} &  \\
		\hline
		J073623.13+392617.7 & [NeV] $\lambda$3427& 0.53\\
		J075217.84+193542.2 &[NeV] $\lambda$3427& 0.26\\
		J090659.46+204810.0 & [FeVII] $\lambda$6086& 3.7\\
		J134918.20+240544.9&  [FeVII] $\lambda$3760 & 1.5\\
		J145420.10+470022.3 & [FeVII] $\lambda$6086& 18.0\\
		J153552.40+575409.4& [FeVII] $\lambda$6086 & 0.81 \\
		J161413.20+260416.3&   [NeV] $\lambda$3347& 0.54  \\
		& [NeV] $\lambda$3427& 0.16 \\
		J171411.63+575834.0 & [NeV] $\lambda$3347&0.10\\
		& [NeV] $\lambda$3427& 2.6\\
		J205141.54+005135.4& [NeV] $\lambda$3427& 0.11\\
		J211646.34+110237.4&  [NeV] $\lambda$3427 & 0.52 \\
		\hline
		\label{tab:residual}
	\end{tabular}
\end{table} 

\cite{ 2018RMxAA..54..217S} also used BPT diagrams to analyze the spectroscopic properties of ionized gas within the central (within 3$^{\prime\prime}$) of the MPL-5 galaxies. The authors similarly used EW H$\alpha$ values to differentiate AGN from post-AGB star ionization regions. However, they relaxed the 3 \AA $\>$ criteria used by \cite{2017MNRAS.472.4382R} and \cite{2010MNRAS.403.1036C}, and only considered galaxies with EW H$\alpha$ $<$ 1.5 \AA $\>$ to be primarily ionized by post-AGB stars (to include weaker AGN). Additionally, the authors excluded galaxies below the Kauffmann demarcation curve, which is an empirical tracer of HII regions - used to isolate star forming sources on BPT diagrams (\citealt{2001ApJ...556..121K, 2006MNRAS.372..961K}). Using these constrains, they reported 98 AGNs in MPL-5. We find that 1/4 CL AGN, that are in MPL-5, are in this catalog (J$2116$). 
  
Finally, \cite{2018MNRAS.474.1499W} analyzed each spaxel for every galaxy in MPL-5 and created spatially resolved BPT-diagrams. They weighted the summed fraction of spaxels  categorized as either AGN, LINER, or composite in each galaxy (e.g., AGN spaxels were given an 80\% weight and composite spaxels were given a 20\% weight). The authors then performed cuts on H$\alpha$ surface brightness and H$\alpha$ EW. H$\alpha$ surface brightness is a reliable tracer of diffuse hot ionized gas; however,  emission line ratios are often enhanced in regions with low H$\alpha$ surface brightness (which can mimic AGN and LINER emission; e.g., \citealt{2009RvMP...81..969H}). As a result, they excluded spaxels with H$\alpha$ surface brightnesses $<$ 10$^{37}$ erg s$^{-1}$ kpc$^{-2}$. They also elevated the minimum EW H$\alpha$ threshold to 5 \AA $\>$ to further reduce potential stellar contamination (below this limit post-AGB stars are considered to be the primary ionization mechanism). They found 303 AGN candidates, and we identify 2/4 CL AGNs, that are in MPL-5, in their sample (J$1535$ and J$2116$). \cite{2018MNRAS.474.1499W} acknowledged that BPT ionization ratios can be impacted by diﬀuse ionized gas, extraplanar gas, photoionization by hot stars, metallicity, and shocks. These sources can elevate line ﬂux ratios to produce AGN-like features, potentially leading to misclassiﬁcation (e.g., \citealt{2010ApJ...721..505R,2011ApJ...734...87R,2013ApJ...774..100K}). 
 
 We determine that the \cite{2017MNRAS.472.4382R}, \cite{ 2018RMxAA..54..217S}, and  \cite{2018MNRAS.474.1499W} catalogs feature 0/4, 1/4, and 2/4 CL AGN(s) (in MPL-5), respectively. The low fraction of CL AGNs in these catalogs suggests that CL detections may be useful for identifying AGNs missed by traditional BPT diagrams.
We will conduct a more thorough review in a forthcoming publication (Negus et al., in prep) and determine the BPT classifications for all of the CL-emitting spaxels in our sample. We will also compare our findings with future MaNGA BPT AGN catalogs that scan a more complete sample of MaNGA galaxies. This will help us build a deeper understanding of the CL galaxies, and will enable us to more closely evaluate the strength of BPT diagrams as AGN classifiers.
\begin{table}[t]
	\renewcommand{\thetable}{\arabic{table}}
	\centering
	\caption{CL Galaxies Found in the MaNGA AGN Catalog}
	\begin{tabular}{cc}
		\hline
		\hline
		SDSS Name & AGN Detection Method \\
		\hline
		J073623.13+392617.7 & \textit{WISE}, Broad\\
		J075217.84+193542.2 &\textit{WISE}, BAT\\
		J153552.40+575409.4 & \textit{WISE}, BAT, Broad\\
		J161413.20+260416.3 & \textit{WISE}, BAT, Broad\\ 
		J171411.63+575834.0 & \textit{WISE}, Broad\\
		J205141.54+005135.4 & \textit{WISE}, Broad\\
		J211646.34+110237.4 & \textit{WISE}, Broad\\
		
		\hline
	\end{tabular}
	\label{tab:mangaagn}
\end{table}  
\newpage
\section{Discussion} \label{sec:discussion}

In this paper, we analyze 10 CL galaxies from MaNGA's MPL-8. With our detection rate (0.16\%), we expect to find at least $\sim$ 16 CL galaxies with CL emission at $\ge$ 5$\sigma$ above the continuum, in at least 10 spaxels, in the final MaNGA catalog of $\sim$ 10,000 galaxies. We anticipate that the majority of these CL galaxies will be confirmed as AGNs or AGN candidates, and that the remaining will be CL mergers.

We determine that AGN photoionization is likely the dominant ionization mechanism for the CLs in our sample, which is consistent with results found in previous studies (e.g., \citealt{2009MNRAS.397..172G, 2010MNRAS.405.1315M}). As a result, we consider CLs to be a useful tracer for AGN identification, which is a critical step in constraining the role of AGN feedback in the host galaxy's evolution. On the other hand, we also determine that CLs can be featured in merging galaxies that may not host an AGN (20\% of our catalog), likely through gas inflows that trigger shocks and extend the reach of the CLR. However, these CL galaxies may still be useful for understanding feedback within galaxies since mergers can initiate galaxy winds, fuel AGNs, stimulate star formation, and impact a galaxy's gas supply (e.g., \citealt{2005MNRAS.361..776S, 2009ApJ...698..956C, 2015ApJS..221...28R}).

Further, \cite{2019BAAA...61..186D} studied CL emission in four nearby AGNs and reported a co-spatial distribution of CL emission with radio jets. \cite{2009MNRAS.394L..16M} also modeled the location and kinematics of the CLR in a sample of AGNs and declared that the bulk of these regions corresponded to outflows. These AGN processes, in addition to shocks, may be additional CL ionization mechanisms that can account for the extended emission we report. 

Finally, prior CL studies measured these lines to lie between the BLR and the NLR (e.g., \citealt{2010MNRAS.405.1315M, 2011ApJ...743..100R}), or on the order of hundreds of pc (e.g.,  \citealt{2011ApJ...739...69M,2021MNRAS.tmp..778R}). Here, we measure the CLR to be far more extended, out to distances of 1.3 - 23 kpc from the galactic center and well into the NLR. We do consider the possibility that the CL emission in the CL galaxies can be smeared (to distances much larger than the FWHM of the PSF) by seeing from the atmosphere, the telescope, and/ or the instruments (e.g., \citealt{2019MNRAS.489..855C}). In Section \ref{sec:results}, we show that beam smearing is possible for 4/10 CL galaxies and consider their CLDs to be upper estimates. However, the remaining population of 6 CL galaxies feature continuous and well resolved CL emission. Among this sample, the CL galaxy with the most extended CL emission is spatially resolved and extends out to 23 kpc. As such, the extent of the CLR that we report (1.3 - 23 kpc from the galactic center) is unlikely to be impacted signiﬁcantly by beam smearing.

\section{Summary and Future Work} \label{sec:conclusion}

We assemble the largest catalog of MaNGA CL galaxies to date. With our custom pipeline, we detect 10 CL galaxies exhibiting emission from one or more CLs ([NeV] $\lambda 3347$, [NeV] $\lambda 3427$, [FeVII] $\lambda 3586$, [FeVII] $\lambda 3760$, or [FeVII] $\lambda 6086$ in this paper) detected at $\ge$ 5$\sigma$ above the background continuum in at least 10 spaxels. 


Our primary results are the following:
\begin{enumerate}
	
\item CL emission extends 1.3 - 23 kpc from the galactic center, with an average distance of 6.6 kpc  (well into the traditional NLR). 

\item Across our entire sample, CL luminosity  diminishes exponentially from the galactic center with -1.8 $\pm$ 0.3 $\le$ $\alpha$ $\le$ 0.2 $\pm$ 0.1. We compare this to the power law index expected for pure AGN photoionization ($\alpha = -2$),  and reason that shocks (e.g., merger-induced and from SNRs) can also produce CLs and increase $\alpha$ to values $>$ -2. 

\item The average CLR electron temperature ranges between 12,331 K - 22,530 K. Shock-induced compression and heating must necessarily elevate these temperatures beyond the threshold for photoionization ($\sim$ 20,000 K). 

\item The average CLR electron number density is on the order of $ \sim 10^{2}$ cm$^{-3}$, consistent with the CLR occupying the NLR, beyond the BLR.

\item CL luminosity strongly correlates with bolometric luminosity (Pearson value of 0.9) for our sample. This is consistent with AGN activity primarily regulating the strength of CLs. 

\item The CL mergers (J$1454$ and J$0906$) deviate most significantly (R$_{f} >$ 3) from the linear regression fit performed on CL luminosity vs. bolometric luminosity.  J$1454$,  8.1$\sigma$ above the mean of the $R_{f}$ distribution, is not a strong AGN candidate based on this result. J$0906$,  1.7$\sigma$ above the mean of the $R_{f}$ distribution,  requires further analysis to determine if it hosts an AGN.

\item 7 CL galaxies (70\% of our catalog) are confirmed AGNs. One CL galaxy is also an AGN candidate. CLs are strong, but perhaps not perfect, indicators of AGN activity. 

\item Several CL AGNs are not found in existing BPT AGN catalogs. Our preliminary results suggests that CL detections may be useful for helping to identify AGNs missed by traditional BPT diagrams.

\end{enumerate} 
We will conduct a full review of the CLR kinematics in a forthcoming publication to determine the role of outflows in the production of CLs. Specifically, we will trace the bulk motion of galactic outflows (i.e. jets), study the dynamics of gas inflows that result from mergers, and measure the rotation and cloud velocities of gas near the galactic core using MaNGA's DAP.

We will also scan MPL-11, the final MaNGA data release that contains $\sim$ 10,000 galaxies, in a follow up investigation and append our sample of CL galaxies with any additional CL galaxy detections. Finally, we will analyze the galaxies with emission from one or more CLs detected at $\ge$ 3$\sigma$ above the background continuum to ensure we identify all CL galaxy candidates and to ultimately build the most complete sample of CL galaxies in MaNGA.



\acknowledgments
\section*{Acknowledgments}
J.N. and J.M.C. acknowledge support from NSF AST-1714503. 
J.N. thanks Mitchell Revalski for useful discussions focused on the physical conditions of ionized gases and emission line diagnostics. RAR acknowledges  partial financial support from CNPq and FAPERGs.

Funding for the Sloan Digital Sky Survey IV has been provided by the Alfred P. Sloan Foundation, the U.S. Department of Energy Office of Science, and the Participating Institutions. SDSS-IV acknowledges
support and resources from the Center for High-Performance Computing at
the University of Utah. The SDSS web site is www.sdss.org.

SDSS-IV is managed by the Astrophysical Research Consortium for the 
Participating Institutions of the SDSS Collaboration including the 
Brazilian Participation Group, the Carnegie Institution for Science, 
Carnegie Mellon University, the Chilean Participation Group, the French Participation Group, Harvard-Smithsonian Center for Astrophysics, 
Instituto de Astrof\'isica de Canarias, The Johns Hopkins University, Kavli Institute for the Physics and Mathematics of the Universe (IPMU) / 
University of Tokyo, the Korean Participation Group, Lawrence Berkeley National Laboratory, 
Leibniz Institut f\"ur Astrophysik Potsdam (AIP),  
Max-Planck-Institut f\"ur Astronomie (MPIA Heidelberg), 
Max-Planck-Institut f\"ur Astrophysik (MPA Garching), 
Max-Planck-Institut f\"ur Extraterrestrische Physik (MPE), 
National Astronomical Observatories of China, New Mexico State University, 
New York University, University of Notre Dame, 
Observat\'ario Nacional / MCTI, The Ohio State University, 
Pennsylvania State University, Shanghai Astronomical Observatory, 
United Kingdom Participation Group,
Universidad Nacional Aut\'onoma de M\'exico, University of Arizona, 
University of Colorado Boulder, University of Oxford, University of Portsmouth, 
University of Utah, University of Virginia, University of Washington, University of Wisconsin, 
Vanderbilt University, and Yale University.
Collaboration Overview
Start Guide
Affiliate Institutions
Key People in SDSS
Collaboration Council
Committee on Inclusiveness
Architects
Survey Science Teams and Working Groups
Code of Conduct
Publication Policy
How to Cite SDSS
External Collaborator Policy

This publication makes use of data products from the Wide-field Infrared Survey Explorer, which is a joint project of the University of California, Los Angeles, and the Jet Propulsion Laboratory/California Institute of Technology, funded by the National Aeronautics and Space Administration.

This research has made use of data supplied by the UK Swift Science Data Centre at the University of Leicester.
 \software{Astropy \citep{2013A&A...558A..33A, 2018AJ....156..123A}}, IRAF \citep{1986SPIE..627..733T, 1993ASPC...52..173T}, PyNeb \citep{2015A&A...573A..42L}.

\newpage
\appendix
\section{Double Gaussian Fit For H$\gamma$ and [OIII] $\lambda4363$}
\label{appendixa}
	\begin{figure}[h]
	\includegraphics[width =\textwidth]{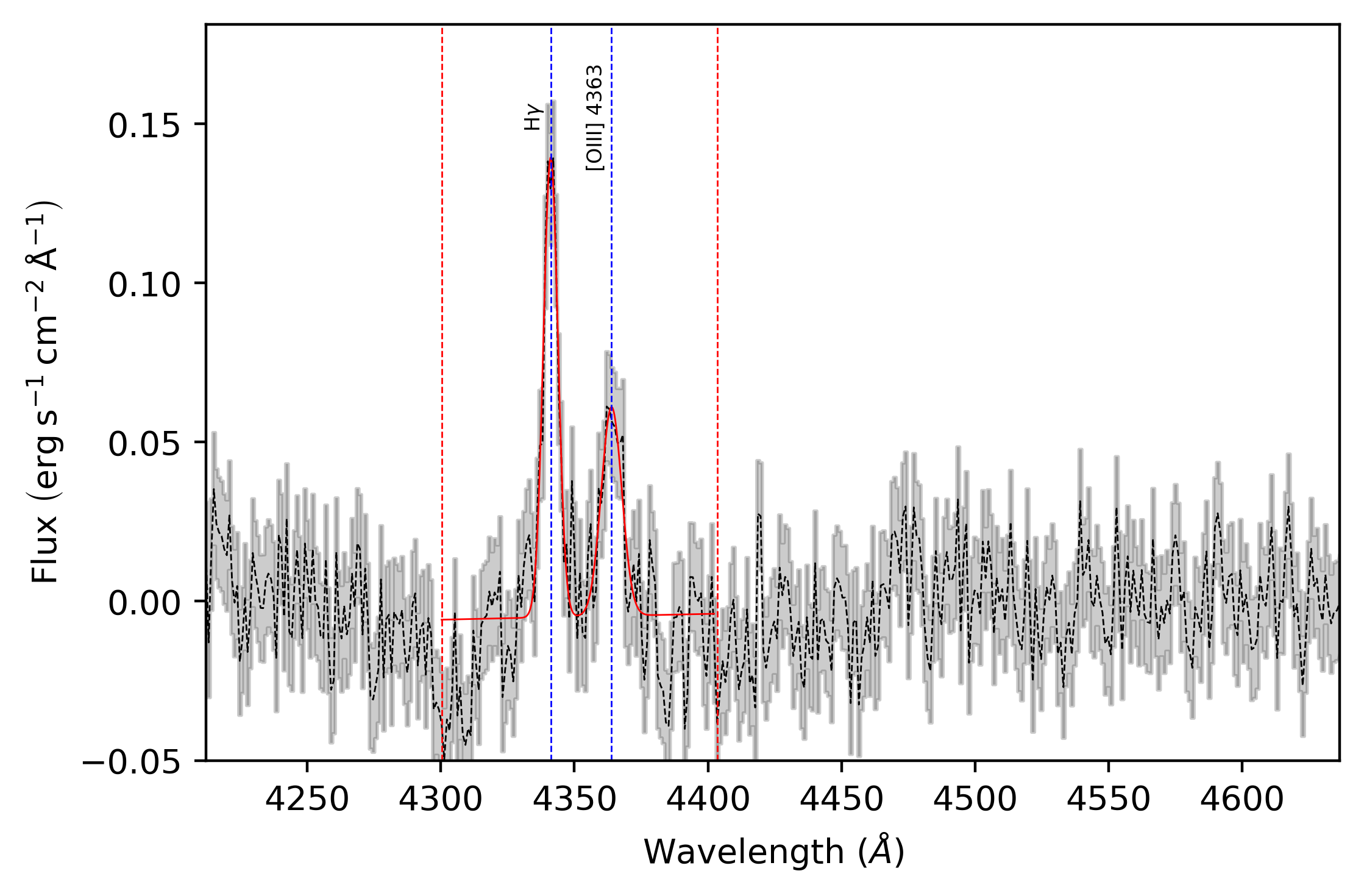}
	\caption{A sample spectrum showing a double Gaussian fit on the H$\gamma$ and the [OIII] $\lambda4363$ lines (detected at $\ge 5 \sigma$ above the continuum) in a single spaxel for J$2116$. The dashed black line is the continuum subtracted spectrum, the shaded gray region is the uncertainty, the solid red line represents the best fit, the red dotted vertical lines mark the fitting window, and the blue dotted lines show the rest wavelengths of H$\gamma$ and [OIII] $\lambda4363$.}
	\label{fig:dg2}
\end{figure}
\newpage
\section{[OIII] $\lambda$5007 Flux Maps for the CL galaxies}
\label{appendixb}

\begin{figure}[h]
	\includegraphics[width=\textwidth,height=5.5cm]{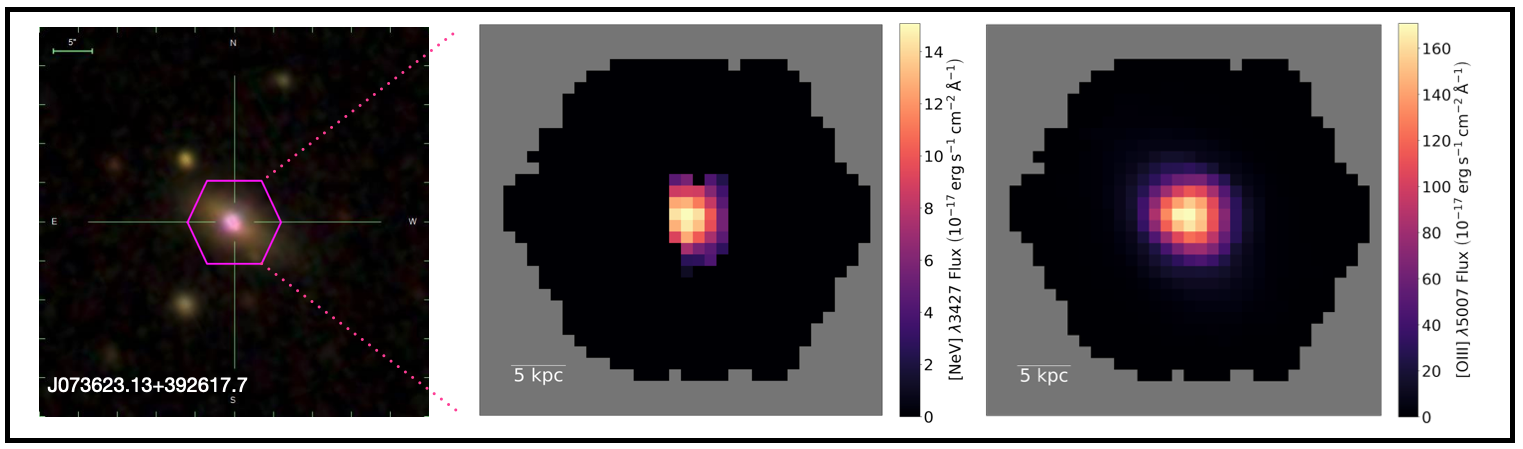}
	\caption{From left to right: SDSS optical image,  [NeV] $\lambda 3427$ flux map, and [OIII] $\lambda5007$ flux map for J$0736$.}
\end{figure}
\begin{figure}[h]
	\includegraphics[width=\textwidth,height=5.5cm]{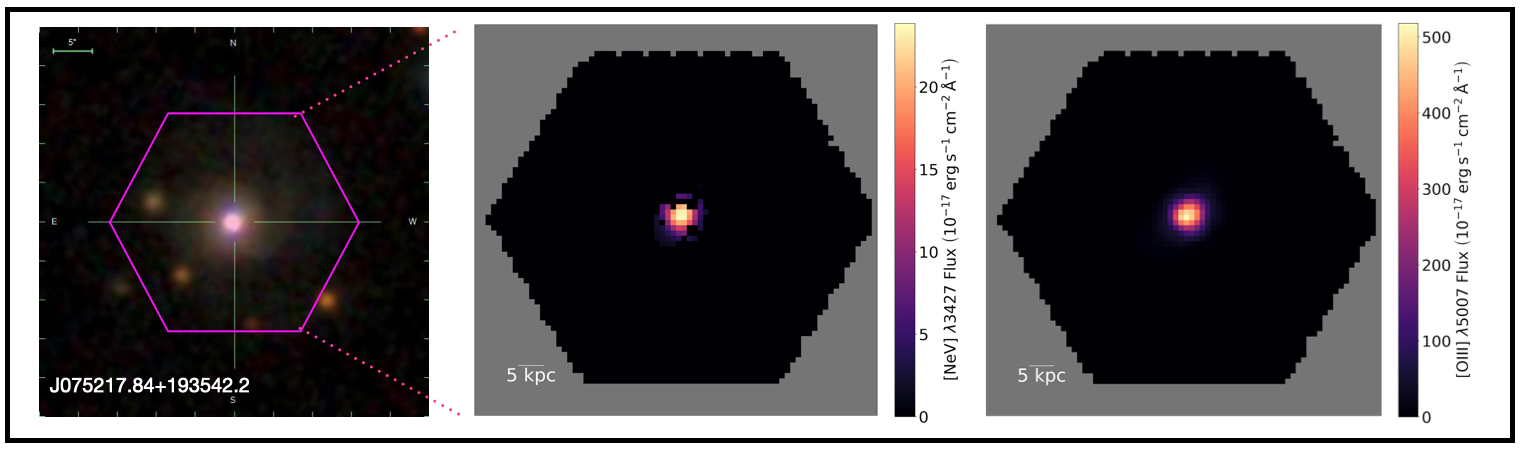}
	\caption{From left to right: SDSS optical image,  [NeV] $\lambda 3427$ flux map, and [OIII] $\lambda5007$ flux map for J$0752$.}
\end{figure}
\begin{figure}[h]
	\includegraphics[width=\textwidth,height=5.5cm]{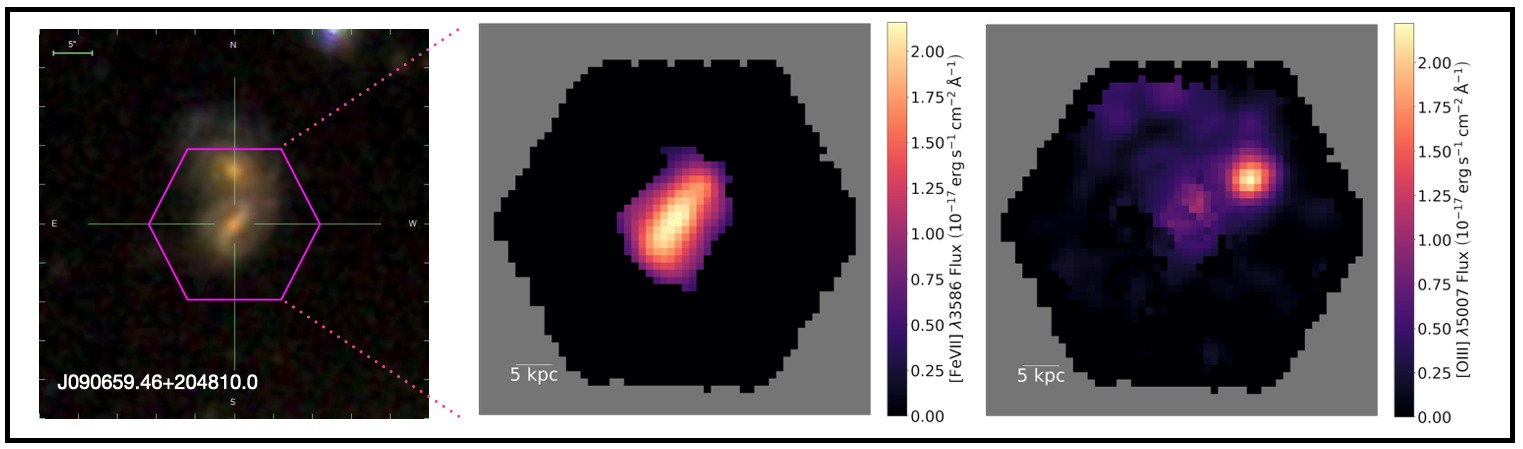}
	\caption{From left to right: SDSS optical image, [FeVII] $\lambda 3586$ flux map, and [OIII] $\lambda5007$ flux map for J$0906$.}
\end{figure}
\begin{figure}[h]
	\includegraphics[width=\textwidth,height=5.5cm]{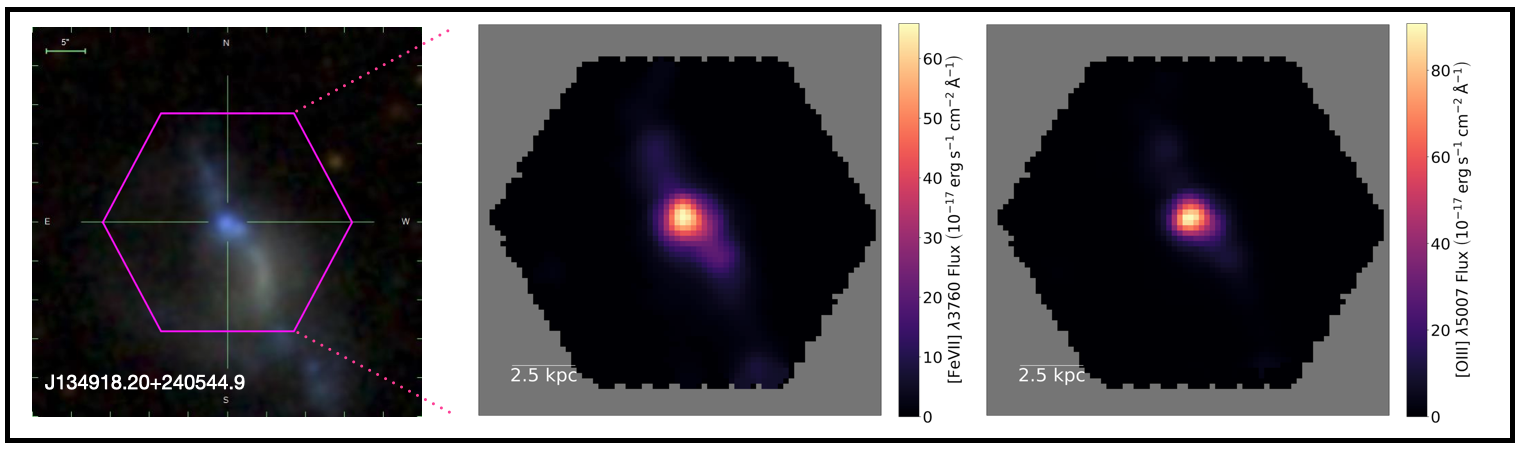}
	\caption{From left to right: SDSS optical image,  [FeVII] $\lambda 3760$ flux map, and [OIII] $\lambda5007$ flux map for J$1349$}
\end{figure}
\begin{figure}[h]
	\includegraphics[width=\textwidth,height=5.5cm]{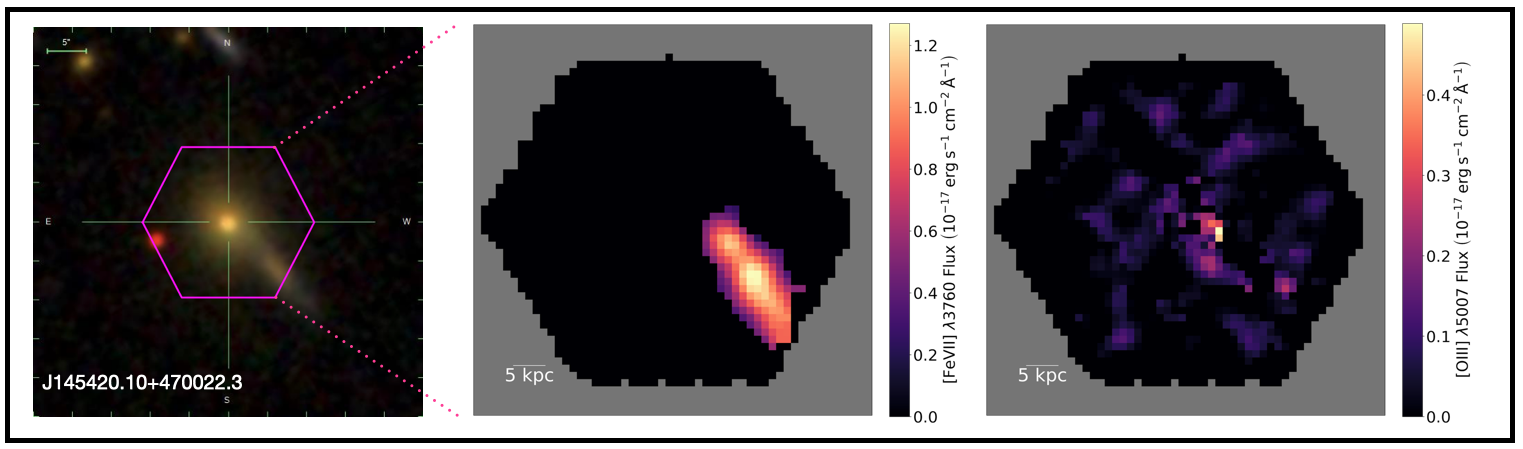}
	\caption{From left to right: SDSS optical image,  [FeVII] $\lambda 3760$ flux map, and [OIII] $\lambda5007$ flux map for J$1454$}
\end{figure}
\begin{figure}[h]
	\includegraphics[width=\textwidth,height=5.5cm]{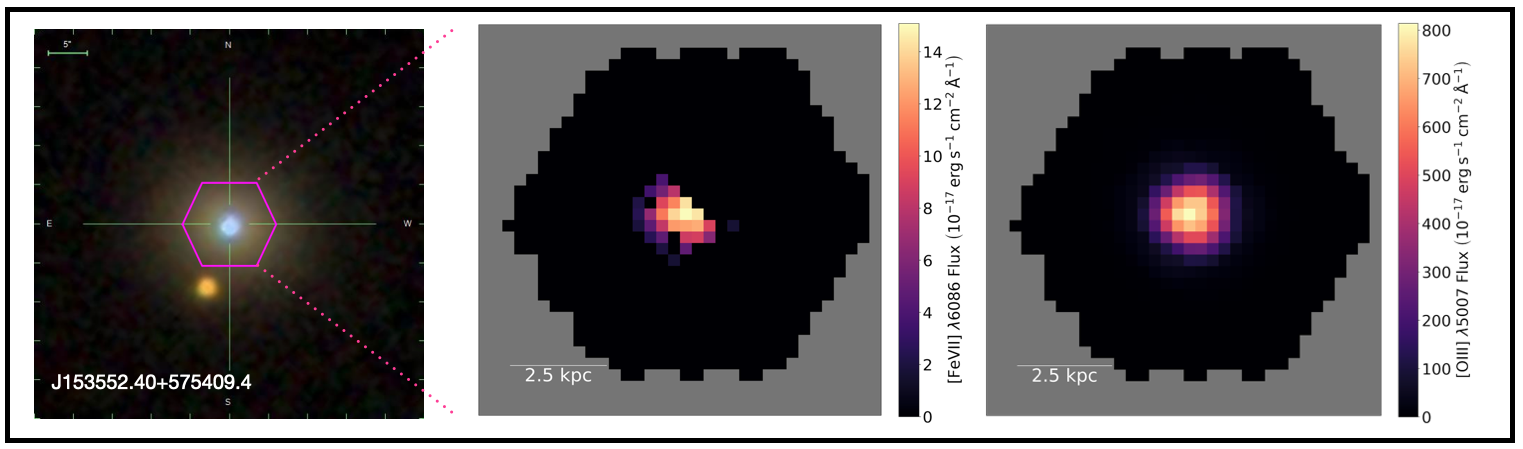}
	\caption{From left to right: SDSS optical image, [FeVII] $\lambda 6086$ flux map, and [OIII] $\lambda5007$ flux map for J$1535$}
\end{figure}
\begin{figure}[h]
	\includegraphics[width=\textwidth,height=15cm]{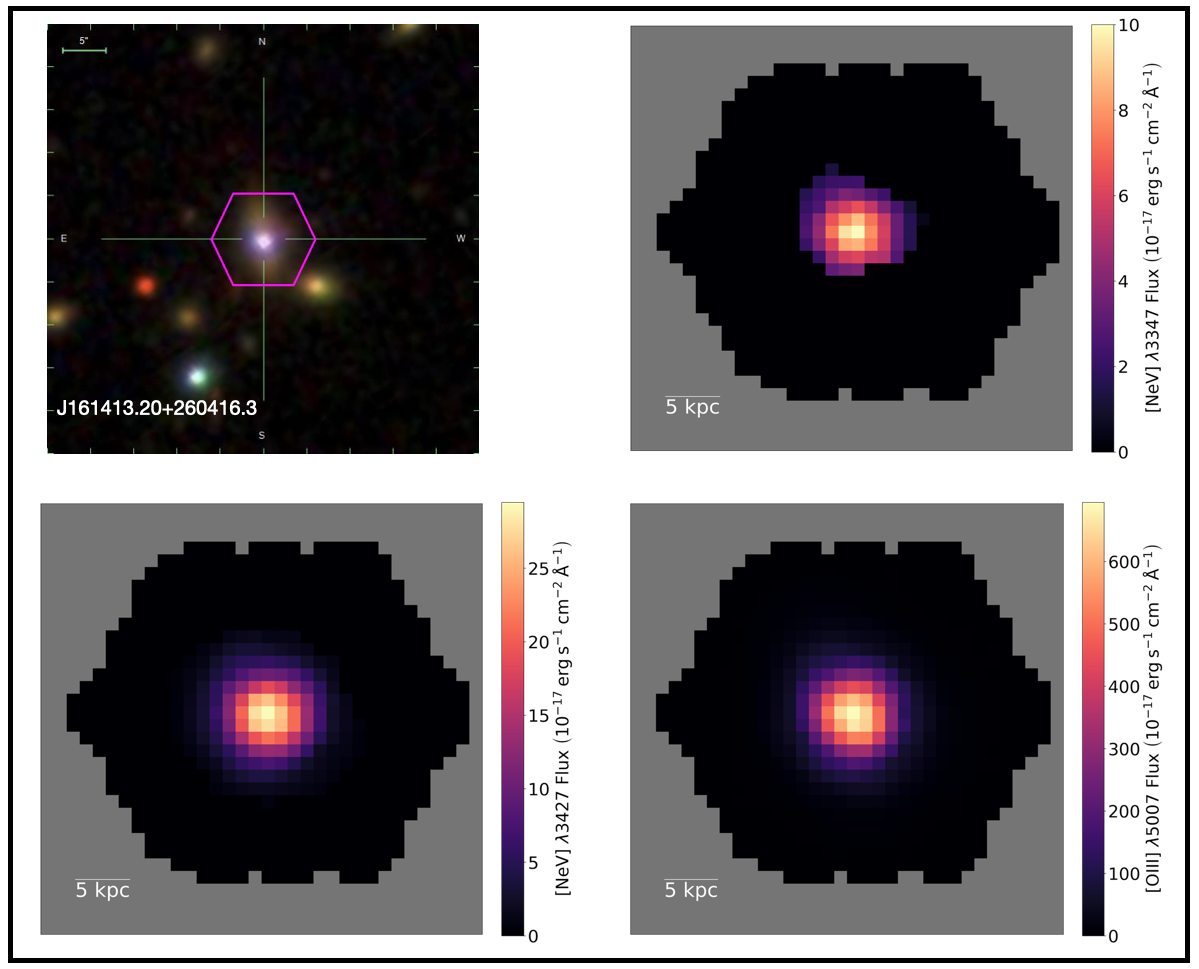}
	\caption{Top row (left to right): SDSS optical image and [NeV] $\lambda 3347$ flux map  for J$1614$. Bottom row (left to right): [NeV] $\lambda 3427$ flux map and [OIII] $\lambda5007$ flux map for J$1614$.}
\end{figure}
\begin{figure}[h]
	\includegraphics[width=\textwidth,height=15cm]{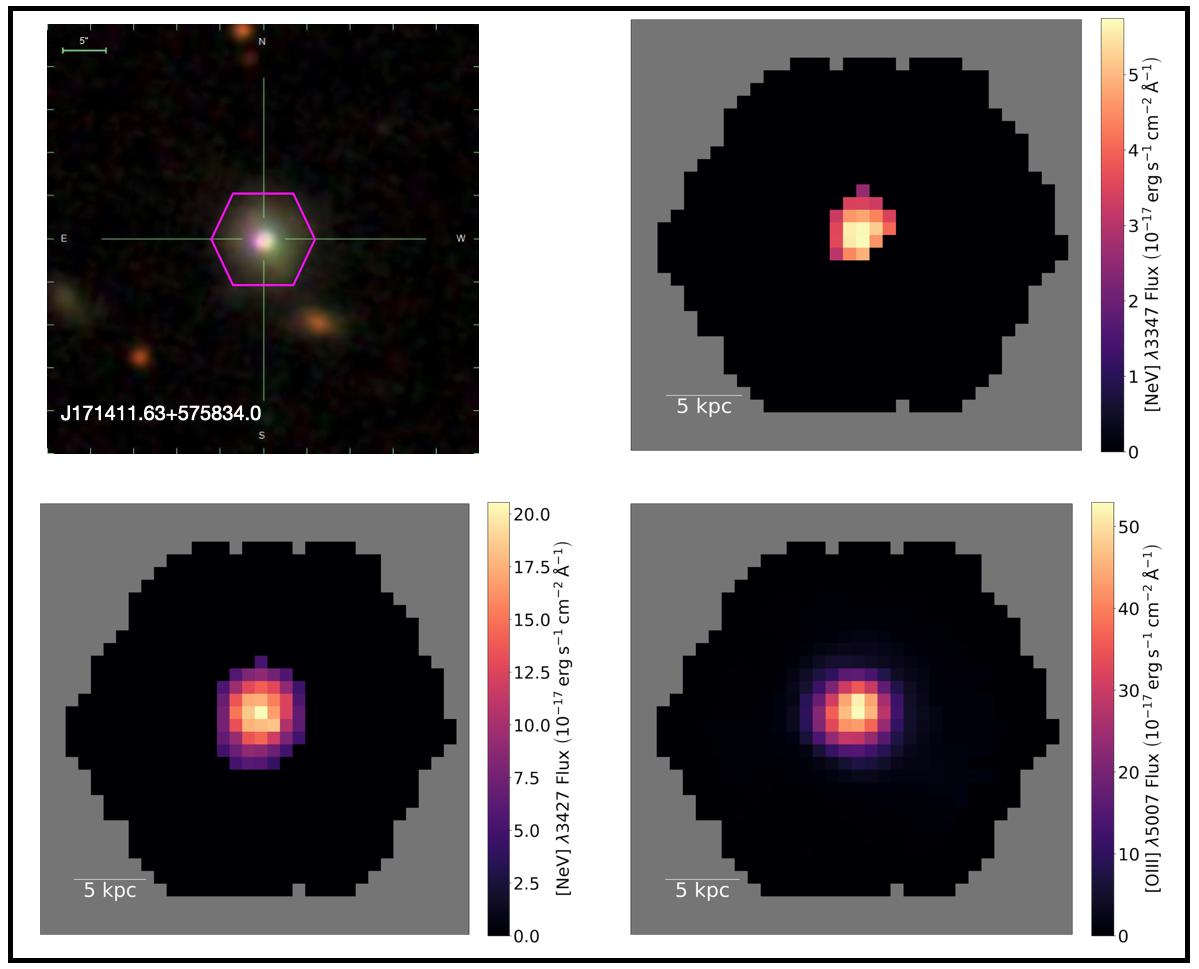}
	\caption{Top row (left to right): SDSS optical image and [NeV] $\lambda 3347$ flux map  for J$1714$. Bottom row (left to right): [NeV] $\lambda 3427$ flux map and [OIII] $\lambda5007$ flux map for J$1714$.}
\end{figure}
\begin{figure}[h]
	\includegraphics[width=\textwidth,height=5.5cm]{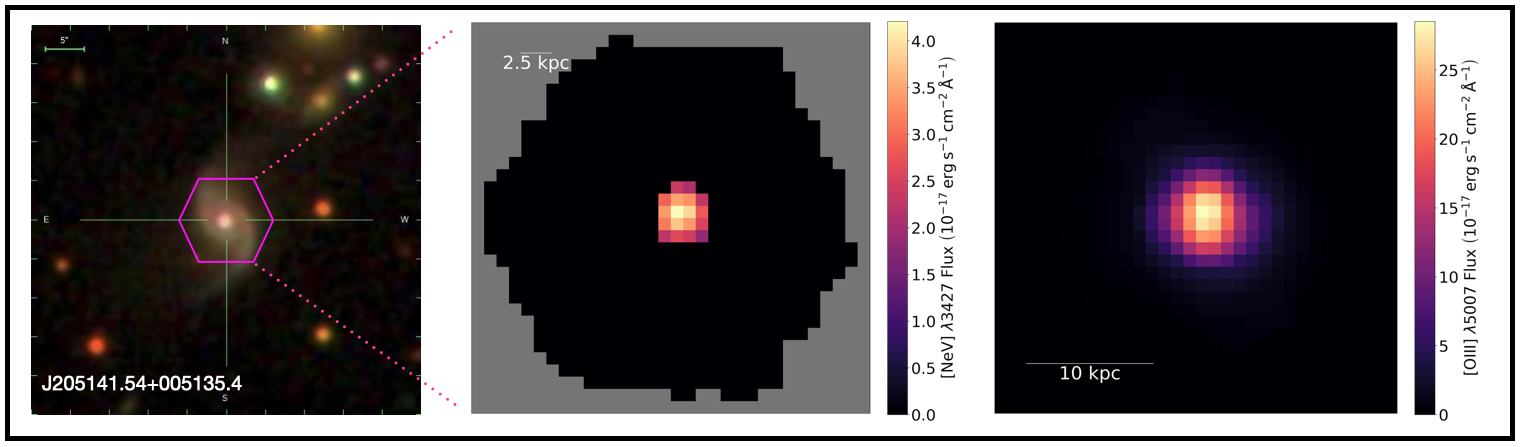}
	\caption {From left to right: SDSS optical image,  [NeV] $\lambda 3427$ flux map, and [OIII] $\lambda5007$ flux map for J$2051$.}
\end{figure}
\begin{figure}[h]
	\includegraphics[width=\textwidth,height=5.5cm]{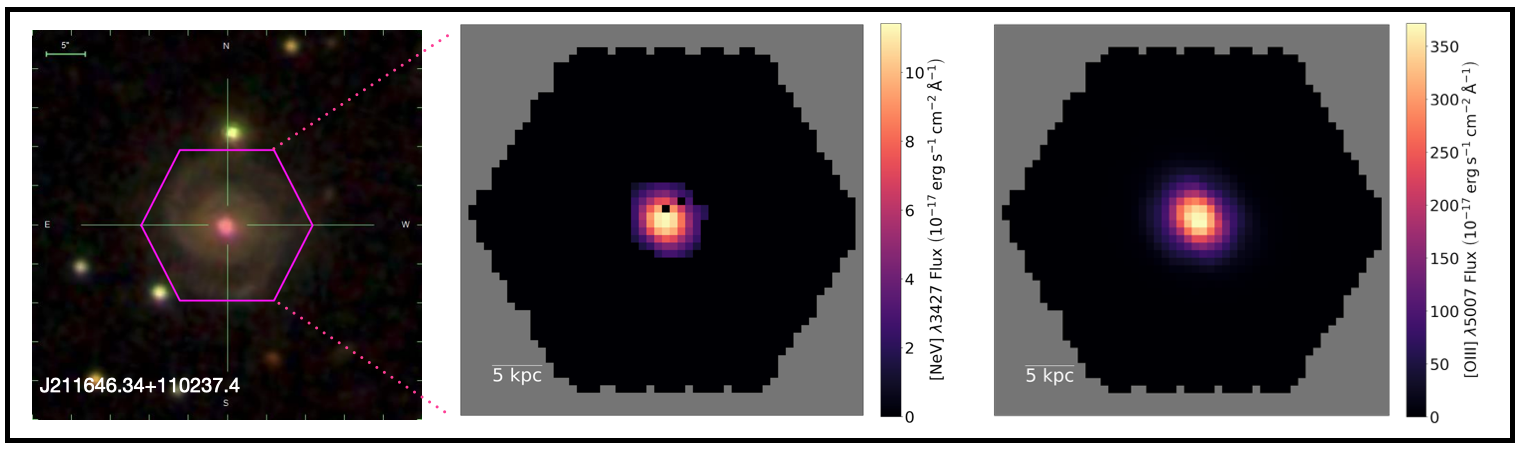}
	\caption{From left to right: SDSS optical image, [NeV] $\lambda 3427$ flux map, and [OIII] $\lambda5007$ flux map for J$2116$.}
\end{figure}

\end{document}